\documentclass[11pt]{article}
\usepackage[pctex32]{graphics}
\usepackage{amssymb}
\usepackage{latexsym}
\usepackage{epsfig}
\usepackage{multirow}
\usepackage{pstcol}
\usepackage{float}
\usepackage{epstopdf}
\usepackage{mathrsfs}

\usepackage{graphicx}
\usepackage{amsmath}
\usepackage{amssymb}
\usepackage{relsize}
\usepackage{multirow}

\usepackage{graphicx}
\usepackage{multicol}

\usepackage{times}
\usepackage{epstopdf}

\definecolor{Red}{rgb}{1,0.,0.}

\usepackage{fancyhdr}
\usepackage{setspace}
\newcommand{\R}{{\mathbb R}}

\newcommand{\bJ}{{\bf J}}

\newcommand{\bX}{{\bf X}}

\newcommand{\bR}{{\bf R}}

\newcommand{\bvarphi}{{\mbox{\boldmath$\varphi$}}}
\newcommand{\bpsi}{{\mbox{\boldmath$\psi$}}}

\newcommand{\bc}{{\bf c}}
\newcommand{\be}{{\bf e}}

\newcommand{\bu}{{\bf u}}
\newcommand{\bv}{{\bf v}}

\newcommand{\bj}{{\bf j}}

\newcommand{\bq}{{\bf q}}

\begin{document}

\title{Energy Demand and Metabolite Partitioning in Spatially Lumped and Distributed Models of Neuron-Astrocyte Complex}
\author{Daniela Calvetti$^1$ \and Yougan Cheng$^2$ \and Erkki Somersalo$^1$}
\date{$^1$Case Western Reserve University \\ Department of Mathematics, Applied Mathematics and Statistics \\ 10900 Euclid Avenue, Cleveland, OH 44106 \\
$^2$University of Minnesota\\ School of Mathematics\\206 Church St SE,
Minneapolis, MN 55455 }
\maketitle
\begin{abstract}
The degrees of freedom of multi-compartment mathematical models for energy metabolism of a neuron-astrocyte complex may offer a key to understand the different ways in which the energetic needs of the brain are met. In this paper we address the problem within a steady state framework and we use the techniques of linear algebra to identify the degrees of freedom first in a lumped model, then in its extension to a spatially distributed case. The interpretation of the degrees of freedom in metabolic terms, more specifically in terms of glucose and oxygen partitioning,  is then leveraged to derive constraints on the free parameters needed to guarantee that the model is energetically feasible. We also demonstrate how the model can be used to estimate the stoichiometric energy needs of the cells as well as the household energy based on observed oxidative cerebral metabolic rate (CMR) of glucose, and the glutamate cycling. Moreover, our analysis shows that in the lumped model the direction of lactate dehydrogenase (LDH) in the cells can be deduced from the glucose partitioning between the compartments. The extension of the lumped model into a spatially distributed multi-compartment setting that includes diffusion fluxes from capillary to tissue increases the number of degrees of freedom, requiring the use of statistical sampling techniques. The analysis of distributed model reveals that some of the conclusions, e.g., concerning the LDH activity and glucose partitioning, based on a spatially lumped model may no longer hold.

keywords: Brain Energy Metabolism \and Bayesian Flux Balance Analysis \and Lactate Shuttle \and Distributed Model \and Glucose Partitioning
\end{abstract}

\section{Introduction}
\label{intro}
Energy metabolism in human brain depends on a complex metabolic network describing the biochemical reactions occurring in the tissue and the regulation of exchanges of metabolic substrates and byproducts between tissue and capillaries. Mathematical models of increasing sophistication have been proposed over the past decade, in response to the difficulty in conducting direct measurements of the quantities of interest in humans without disrupting the brain functions: see, e.g.,   \cite{Aubert2005,Calvetti2011,Cloutier,Dinuzzo,Simpson}. Some of these models assume that the brain is in a steady or a stationary state, where either the concentrations of the metabolites and intermediates or their derivatives are constant, and look for a configuration of reaction fluxes and cross-membrane transport rates which can maintain the brain in a given state. In these models, which are governed by systems of linear equations with constrains on some of the unknowns, the attention is on the activity level of the various reactions and transports, and no information about the concentrations of the biochemical species involved is provided. Kinetic metabolic models, on the other hand, whose aim is to describe the time courses of the different metabolites, are governed by systems of differential equations depending on a large number of parameters whose values are either unknown or poorly known.  A standard method for estimating these unknown model parameters is to fit the model predictions to measured concentration data using a reliable numerical method. The underlying parameter estimation problem for detailed models which depend on a large number of unknown parameters may be challenging because of the limited amount of available data, which, in addition, often come from a cohort of subjects, thus introducing an intrinsic model discrepancy.  A possible alternative would be to assign the parameters values deduced indirectly from in vitro measurements of quantities which may include transporter density on membranes or enzyme expression levels. In addition to the difficulty of obtaining such measurements, the passage from this kind of  data to model parameters is not at all straightforward, in particular, because the connection between the cellular level measurements and the lumped macroscopic model is poorly understood \cite{Calvetti2014}.

In this paper  we start with a reduced lumped model of brain cellar metabolism and propose an interpretation in metabolic terms of the degrees of freedom not determined by the underdetermined model. In particular, we demonstrate that glucose and oxygen partitioning between neuron and astrocyte in the simple lumped model can be interpreted as the free parameters that determine completely the steady state. In particular, when the household energy of the cells is fixed, one can derive a linear relation between the oxidative cerebral metabolic rate (CMR) of glucose and the neurotransmitter cycling activity. Such relation has been empirically verified, and therefore the formula provides a means to estimate the various parameters of the energetics of the neuron-astrocyte complex. We discuss this observation in the light of published measured values as well as electrophysiology-based cell level models.

Subsequently, we show that when moving from a spatially lumped to a distributed model, the number of degrees of freedom increases, and the model becomes analytically intractable. To analyze the spatially distributed model, we modify a computational statistical method previously proposed for the analysis of complex, lumped, metabolic models. The sampling-based statistical methods reveal that conclusions based on the lumped model, while enlightening in many ways, may be too simplistic due to the reduced dimensionality of the system.

\section{Single Unit: A Simplified Neuron-Astrocyte Complex}
\label{sec:single unit}
In this section, we consider a simplified, spatially lumped compartment model for neuron-astrocyte interactions during neurotransmitter cycling  whose nontrivial null space can be a described explicitly and interpreted in metabolic terms. In particular, we relate the two degrees of freedom in the model to the glucose and oxygen partitioning between the neuron and the astrocyte, and show how they affect the lactate shuttling between the cells. In order for the model to be physiologically meaningful, the metabolic flux configuration must be able to meet the aggregate energetic needs of the complex, which comprise the cost for performing basic household chores and the energy required to support the signaling activities. This in turn translates into a set of constraints for the reaction fluxes and transport rates, which will be derived starting from the underlying physiological motivations.

\subsection{Stoichiometry and Degrees of Freedom}
\label{subsec:stoichiometry}
Consider a simplified mathematical model of a neuron-astrocyte complex, equipped with the transports of glucose, lactate, oxygen and carbon dioxide between blood, extracellular space (ECS) and the cell compartments comprising, in each cell type, a lumped glycolysis, lactate dehydrogenase (LDH), integrated pyruvate dehydrogenase (PDH) and tricarboxylic acid (TCA) cycle, and oxidative phosphorylation (OxPhos). Furthermore, we assume that the two cell types are coupled through a glutamate-glutamine cycling over the synaptic cleft: we refer to the neurotransmitter cycle as the V-cycle.

The energetic needs for keeping the glutamatergic signaling active at steady rate are different for neuron and astrocyte. In the neuron, the energetic cost can be broken up in three main categories: (a) Presynaptic activities including sodium-calcium exchange and glutamate packing in vesicles; (b) The post-synaptic cost consisting mostly of the energy needed for the sodium-potassium pumps as well as sodium-calcium exchange; (c) Action potential propagation to trigger synaptic signaling where energy is required for depolarization of the membrane. In astrocyte, energy is needed for glutamate uptake, glutamine synthetase as well as glutamine packing to vesicles. We refer to \cite{Attwell2001} for more detailed discussion of the mechanisms.

\begin{figure*}
\centerline{\includegraphics[width=12cm]{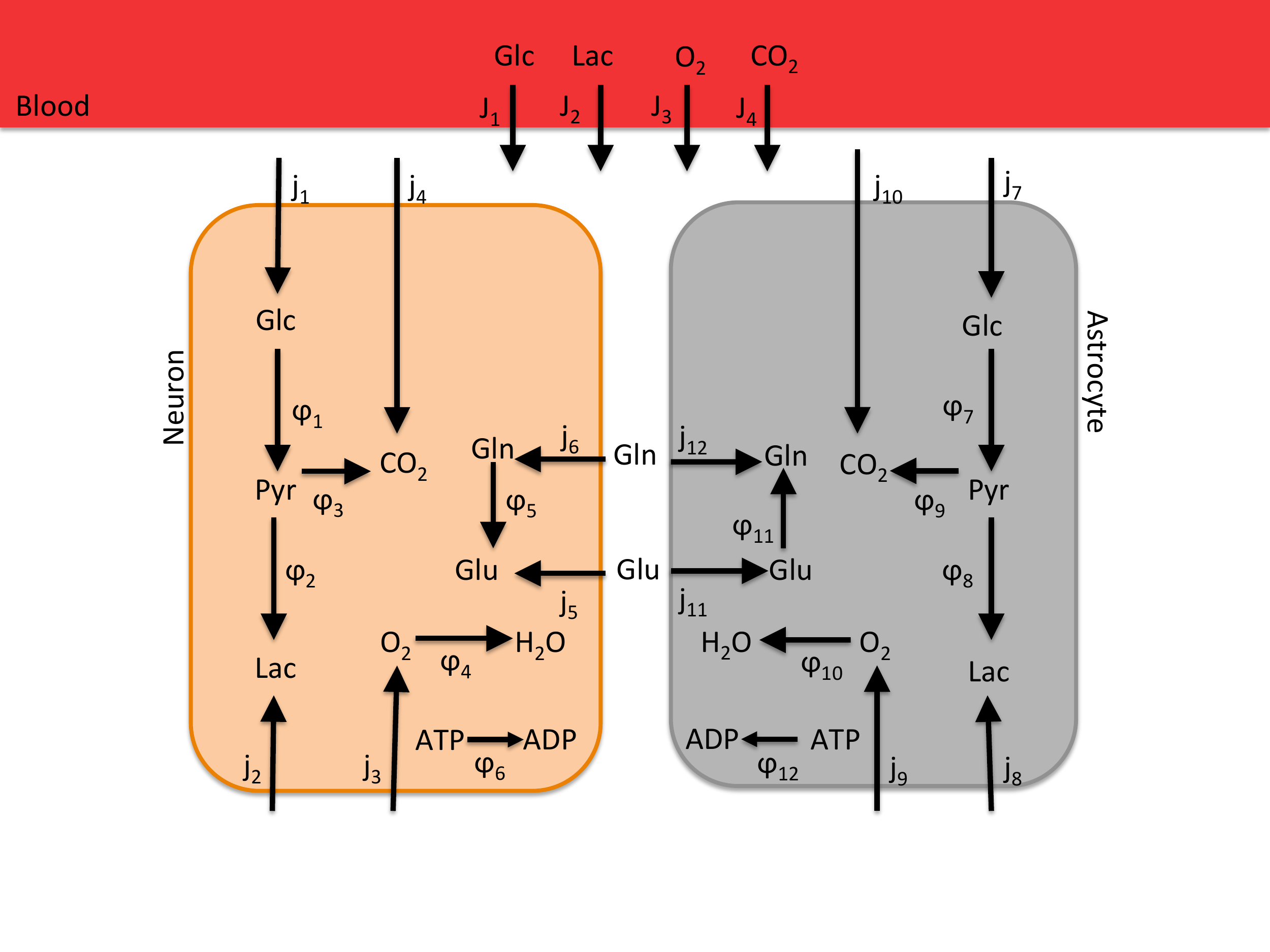}}
\caption{\label{fig:small} The pathway chart and compartment configuration of the  model considered in the simplified metabolic network. The arrows in the diagram indicate the direction of the flux with a positive sign, which may not be the preferred direction. The ATP/ADP and NADH/NAD$^+$ dynamics are not indicated in the figure. }
\end{figure*}

To include the energetic cost in the model, we let $E_{\rm n}$ and $E_{\rm a}$ denote the stoichiometric energy cost in neuron and astrocyte, respectively. More precisely, $E_{\rm n}$ ($E_{\rm a}$)  as the is number of ATP molecules needed in neuron (astrocyte) to cycle one molecule of glutamate through the V-cycle. We will discuss the values of $E_{\rm n}$ and $E_{\rm a}$ in detail later on. Recall that the hydrolysis of ATP releases $\Delta E = 30.5\;{\rm  kJ}/{\rm mol}$ of energy. To implement the energetic cost into the model in a stoichiometric manner, to avoid complicating the model significantly, e.g., by including the transmebrane ion fluxes as new unknownswe, we attach the energy consumption of the transmitter cycle with the V-cycle reactions phosphate activated glutaminase (PAG) in neuron and glutamine synthetase (GS) in astrocyte. The lumped reactions included in this model are listed in Table~\ref{tab:reactions}, and the metabolic network is illustrated schematically in Figure~\ref{fig:small}.  Observe that in Table~\ref{tab:reactions}, we list the ATP hydrolysis, with fluxes $\varphi_6$ and $\varphi_{12}$ in neuron and astrocyte, as a separate reaction. These fluxes are related to the household energetic needs of the cells, which are incurred independently of the V-cycle activity level. The household energy and its implementation into the model will be discussed in detail later on.

\begin{table}
\caption{\label{tab:reactions} List of the lumped reactions included in the model. The phosphate activated glutaminase (PAG), or $\varphi_5$, accounts for the total neuronal energetic need required to maintain the steady state glutamine/glutamate cycle. The energetic need $E_n$ will be specified in the computed examples. Similarly, $\varphi_{11}$ in astrocyte, representing the glutamine synthetase (GS), accounts for the astrocytic energetic need, defined by $E_a$. The ATP dehydrogenase reactions, $\varphi_{6}$ and $\varphi_{12}$, account for non-specific energy consuming processes in the cells. All the reactions except for lactate dehydrogenase (LDH), denoted by $\varphi_2$ in neuron and $\varphi_7$ in astrocyte, are unidirectional.}
\begin{center}
\begin{tabular}{ccl}
Neuron & Astrocyte & Reaction \\
\hline
$\varphi_1$ & $\varphi_7$     &   ${\rm Glc} + 2 \,{\rm NAD}^+ + 2\,{\rm ADP}  \longrightarrow    2\,{\rm Pyr} + 2\, {\rm NADH} + 2\,{\rm ATP}$ \\
$\varphi_2$ & $\varphi_8$     &  ${\rm Pyr} + {\rm NADH}                                 \longrightarrow   {\rm Lac} +{\rm NAD}^+$ \\
$\varphi_3$ & $\varphi_9$     &  ${\rm Pyr} + {\rm ADP} + 5\, {\rm NAD}^+     \longrightarrow    3\,{\rm CO}_2 +{\rm ATP} + 5\,{\rm NADH}$ \\
$\varphi_4$ & $\varphi_{10}$     &  ${\rm O}_2 + 2{\rm NADH} + 5{\rm ADP}     \longrightarrow    2\,{\rm NAD}^+ + 5\,{\rm ATP} + 2\,{\rm H}_2{\rm O}$ \\
$\varphi_5$ &         --                & ${\rm Gln} + E_n\,{\rm ATP} \longrightarrow {\rm Glu} +E_n\,{\rm ADP}$ \\
     --               & $\varphi_{11}$ &${\rm Glu} + E_a\,{\rm ATP} \longrightarrow {\rm Gln} +E_n\,{\rm ADP}$ \\
$\varphi_{6}$ & $\varphi_{12}$ & ${\rm ATP}\rightarrow{\rm ADP}$ \\
\hline
\end{tabular}
\end{center}
\end{table}

\begin{table}
\caption{\label{tab:transports} Transport fluxes from extracellular space (ECS) to neuron (n) or astrocyte (a). According to the sign convention, a positive flux is from ECS to the cell. Observe that because of physiological considerations, some of the fluxes are constrained to be positive ($j_1, j_3, j_6, j_7,j_9, j_{11}$), and some negative ($j_4,j_5,j_{10},j_{12}$). The table includes the stoichiometric connections between the reaction and transport fluxes.}
\begin{center}
\begin{tabular}{ll|ll}
\multicolumn{2}{c}{Neuron} & \multicolumn{2}{c}{Astrocyte} \\
\hline
Transport & Stoichiometry & Transport & Stoihiometry \\
\hline
$j_1: {\rm Glc}_{\rm ECS}\rightarrow{\rm Glc}_{\rm n}$ & $j_1 = \varphi_1$  & $j_7: {\rm Glc}_{\rm ECS}\rightarrow{\rm Glc}_{\rm a}$ & $j_7 = \varphi_7$ \\
$j_2: {\rm Lac}_{\rm ECS}\rightarrow{\rm Lac}_{\rm n}$ & $j_2 = -\varphi_2$  & $j_8: {\rm Lac}_{\rm ECS}\rightarrow{\rm Lac}_{\rm a}$ & $j_8 = -\varphi_8$ \\
$j_3: {\rm O}_{2,{\rm ECS}}\rightarrow{\rm O}_{2,{\rm n}}$ & $j_3 = \varphi_4$  & $j_9: {\rm O}_{2,{\rm ECS}}\rightarrow{\rm O}_{2,{\rm a}}$ & $j_9 = \varphi_{10}$ \\
$j_4: {\rm CO}_{2,{\rm ECS}}\rightarrow{\rm CO}_{2,{\rm n}}$ & $j_4 = -3\,\varphi_3$  & $j_{10}: {\rm CO}_{2,{\rm ECS}}\rightarrow{\rm CO}_{2,{\rm a}}$ & $j_{10} = -3\,\varphi_9$ \\
$j_5: {\rm Glu}_{\rm ECS}\rightarrow{\rm Glu}_{\rm n}$ & $j_5 = -\varphi_5$  & $j_{11}: {\rm Glu}_{\rm ECS}\rightarrow{\rm Glu}_{\rm a}$ & $j_{11} = \varphi_{11}$ \\
$j_6: {\rm Gln}_{\rm ECS}\rightarrow{\rm Gln}_{\rm n}$ & $j_6 = \varphi_5$  & $j_{12}: {\rm Gln}_{\rm ECS}\rightarrow{\rm Gln}_{\rm a}$ & $j_{12} = -\varphi_{11}$ \\
\hline
\end{tabular}
\end{center}
\end{table}

\begin{table}
\caption{\label{tab:stoichiometry} Stoichiometric relations between the reaction and transport rates that follow from balancing the concentrations of the indicated species in each compartment.}
\begin{center}
\begin{tabular}{llll}
Species & Neuron & Astrocyte & ECS \\
\hline
Pyr  & $2\,\varphi_1 - \varphi_2-\varphi_3 = 0$ & $2\,\varphi_7 - \varphi_8-\varphi_9 = 0$ & -- \\
ATP & $2\,\varphi_1 +\varphi_3 + 5\,\varphi_4 -E_{\rm n}\,\varphi_5 -\varphi_{6} =0$ &  $2\,\varphi_7 +\varphi_9 + 5\,\varphi_{10} -E_{\rm a}\,\varphi_{11} -\varphi_{12} =0$ & -- \\
NADH & $2\,\varphi_1 - \varphi_2 + 5\,\varphi_3 - 2\,\varphi_4 =0$ & $2\,\varphi_7  -\varphi_8 + 5\,\varphi_9 - 2\,\varphi_{10} =0$ &  --\\
Glu & (see Table~\ref{tab:transports}) & (see Table~\ref{tab:transports}) &$j_5 = -j_{11}$ \\
Gln & (see Table~\ref{tab:transports}) & (see Table~\ref{tab:transports}) & $j_6 = -j_{12}$ \\
\hline
\end{tabular}
\end{center}
\end{table}

To set up the model, we denote by $\bJ = [J_1;J_2;J_3]\in\R^3$ the column vector of the aggregated tissue uptake of glucose, lactate and oxygen, respectively, from the capillary blood, with the sign convention that $J_j>0$ indicates transport into the tissue. The ratio of oxygen to glucose uptake, referred to as the oxygen glucose index, OGI, is defined by
\[
 {\rm OGI} = \frac{J_3}{J_1}.
\]
It follows from the stoichiometry that the complete oxidation of glucose in tissue,
\[
 {\rm C}_6 {\rm H}_{12} {\rm O}_6+6\, {\rm O}_{\rm 2} \rightarrow  6\,{\rm CO}_2 + 6\,{\rm H}_2{\rm O},
\]
 requires that ${\rm OGI} = 6$: an OGI smaller than 6 indicates the presence of non-oxidative metabolism, or production of lactate, while an OGI greater than 6 signals the oxidation of substrates other than glucose, which in the present model are limited to lactate. More generally, the balance equation including the lactic acid is
 \begin{equation}\label{blood stoichiometry}
 {\rm C}_6 {\rm H}_{12} {\rm O}_6+{\rm OGI}\times {\rm O}_{\rm 2} \rightarrow \left(2-\frac{{\rm OGI}}3\right)\times{\rm C}_3{\rm H}_6{\rm O}_3 + {\rm OGI}\times{\rm CO}_2 + {\rm OGI}\times{\rm H}_2{\rm O}.
\end{equation}
We remark that the coefficient of lactic acid may become negative when ${\rm OGI}>6$, indicating uptake, rather than release, of this metabolite.

Next we will show that the redox balance in the tissue implies that $J_4 = -J_3$, as already indicated in (\ref{blood stoichiometry}).
Consider the reductive/oxidative reactions $\varphi_1,\ldots,\varphi_4$ in the neuron, listed in Table~\ref{tab:reactions}.
By subtracting side by side the pyruvate balance equation
\begin{equation}\label{pyruvatebalance}
2\,\varphi_1 -\varphi_2 - \varphi_3 =0
\end{equation}
and the NADH balance equation
\begin{equation}\label{NADHbalance}
 2\, \varphi_1-\varphi_2 + 5\, \varphi_3 - 2\,\varphi_4 =0,
 \end{equation}
we obtain
\begin{equation}\label{oxygen}
 6\,\varphi_3 -2\,\varphi_4 = 0\quad\mbox{or}\quad 3\,\varphi_3 = \varphi_4.
\end{equation}
Solving for $\varphi_3$ and substituting the result in the pyruvate balance equation yields
\begin{equation}\label{balance}
6\,\varphi_1 - 3\,\varphi_2 -\varphi_4 = 0.
\end{equation}
Repeating the same procedure for astrocyte, we have that
\begin{equation}\label{balance 2}
6\,\varphi_6 - 3\,\varphi_7 -\varphi_9 = 0.
\end{equation}

This last equality, together with the stoichiometric relations $j_3=\varphi_4$ and $j_4 = -3\,\varphi_3$, see Table~\ref{tab:transports}, implies that $j_4 = -j_3$, that is, the carbon dioxide efflux rate equals that of oxygen influx in neuron. The same holds in the astrocyte, thus implying that $J_4 = -J_3$.  In the light of these observations, it suffices to consider the transfer of glucose, lactate, and oxygen.

Let $\bj^{\rm n} = [j_1;j_2;j_3]$ be the column vector whose entries are the uptake rates of glucose, lactate and oxygen, respectively, in neuron,  and $\bj^{\rm a}=[j_7;j_8;j_9]$ the corresponding rate vector in astrocyte. From the convention that a positive flux goes from the ECS into the cell and the fact that at steady state, the equalities $ [j_1;j_2;j_3] = [\varphi_1;-\varphi_2;\varphi_4]$ and $ [j_7;j_8;j_9] = [\varphi_6;-\varphi_7;\varphi_9]$  must hold, it follows from (\ref{balance}) and (\ref{balance 2}) that
\begin{equation}\label{orthogonality 0}
 6\,j_1+3\,j_2 - j_3 = 0,\quad 6\,j_7+3\,j_8 - j_9 = 0.
\end{equation}
Collecting the coefficients of the transport rates into the vector $  \bq = [6;3;-1]$, condition (\ref{orthogonality 0}) can be recast in terms of orthogonallity between pairs of vectors, i.e.,
\begin{equation}\label{orthogonality}
 \bq^{\mathsf T} \bj^{\rm n} = \bq^{\mathsf T} \bj^{\rm a} = 0,
\end{equation}
where the superscript ``${\mathsf T}$'' indicates transposition. It follows from the conservation of flux, $  \bJ = \bj^{\rm n}+\bj^{\rm a},$ that
\begin{equation}
\label{Total sum}
 \bq^{\mathsf T} \bJ = 0.
\end{equation}
This implies that the three-vectors $\bj^{\rm a}$, $\bj^{\rm n}$ , $\bJ$  $\in \mathbb{R}^3$ belong to the plane orthogonal to the vector $\bq$, thus they can be expressed as a linear combination of any two linearly independent vectors $\bv_1$ and $\bv_2$, orthogonal to $\bq$.
Assuming that $\bJ$ is given and writing
\begin{equation}\label{interpret}
\frac{1}{2}(\bj^n-\bj^a)=\alpha_1\bv_1+\alpha_2\bv_2,
\end{equation}
for some scalars $\alpha_1$ and $\alpha_2$, it follows that the general solution $(\bj^{\rm n},\bj^{\rm a})$ of (\ref{orthogonality}) can be expressed in the form
\begin{equation}
\label{circle}
 \bj^{\rm n} = \alpha_1 \bv_1 + \alpha_2 \bv_2 + \frac 12 \bJ,\quad \bj^{\rm a} = -\alpha_1 \bv_1 - \alpha_2 \bv_2 + \frac 12 \bJ.
\end{equation}
Because the vectors $\bv_1$ and $\bv_2$ can be chosen arbitrarily in the plane orthogonal to $\bq$, we conclude that the present model has two degrees of freedom.

To give an interpretation to the degrees of freedom in metabolic terms, we choose the vectors $\bv_1$ and $\bv_2$ to be
\[
 \bv_1 = \left[\begin{array}{r} 1 \\ -2 \\ 0\end{array}\right],\quad \bv_2 = \left[\begin{array}{c} 0\\ 1/3 \\ 1\end{array}\right].
\]
It follows from this choice of the basis vectors and (\ref{interpret})  that $\alpha_1$ controls the difference in glucose uptake between neuron and astrocyte, and $\alpha_2$ the difference in oxygen uptake.

So far, no constraints have been imposed on the vectors $\bj^n$ and $\bj^a$ to guarantee that the system is in an physiologically meaningful state and its  activity level is energetically sustainable. At steady state, it is reasonable to assume that glucose and oxygen can only be taken up, not released, by the cells. This is guaranteed if we require the components $j_1$, $j_3$, and $j_7$, $j_9$ to be non-negative,
\begin{equation}\label{j positive}
 j_1\geq 0,\quad j_3\geq 0,\quad j_7\geq 0,\quad j_9\geq 0.
\end{equation}
With our current choice of $\bv_1$ and $\bv_2$, these conditions translate into constraints for coefficients $\alpha_1$ and $\alpha_2$, namely
\[
 -\frac 12 J_1\leq \alpha_1  \leq \frac 12 J_1,
\]
\[
 -\frac 1{2} J_3\leq \alpha_2\leq \frac 1{2} J_3.
\]
The constrains needed to ensure energetic sustainability will be discussed next.
\subsection{Energy Estimates}

 To guarantee that the metabolic model is energetically feasible, the ATP production must be sufficient to meet the energetic needs of the cells.
 The total ATP production rate in neuron,
\[
 \Phi_{\rm ATP}^{\rm n} = 2\,\varphi_1 +  \varphi_3 +5\,\varphi_4= 2\,\varphi_1 + \left(\frac 13 + 5\right)\varphi_4,
\]
which follows from (\ref{oxygen}), can be expressed in terms of the metabolite transport fluxes as
\begin{equation}\label{ATP_n}
 \Phi_{\rm ATP}^{\rm n} = 2\,j_1 +  \frac {16}3\, j_3,
\end{equation}
and similarly, in astrocyte
\begin{equation}\label{ATP_a}
 \Phi_{\rm ATP}^{\rm a} = 2\,j_7 +  \frac {16}3 \, j_9.
\end{equation}
In order for the metabolic state of the neuron to be energetically sustainable, the ATP production needs to be sufficient for both the neuron signaling and the household costs.

Let $V$ denote the steady state neurotransmission activity level of the system, measured in terms of the V-cycle rate,
\[
 \varphi_5 = \varphi_{10} = V.
\]
In addition to neurotransmission and signaling, the cells need energy for other routine tasks, e.g., for maintaining the membrane potentials. We refer to the latter energy demand as household energy, and denote it by $H_{\rm n}$ and $H_{\rm a}$ in neuron and in astrocyte, respectively.  Rather than complicating the model with a detailed description of the underlying electrophysiology, we require that part of the ATP hydrolysis takes place to provide the household energy, thus
\begin{equation}\label{household condition}
 \varphi_6 \geq  H_{\rm n}, \quad \varphi_{12} \geq H_{\rm a}.
\end{equation}
To guarantee that the ATP turnover in neuron and astrocyte is enough to maintain the activity level of the system, we require that it suffices to meet the energetic requirements of the household tasks and neurotransmission,  as expressed by the inequalities
\begin{equation}\label{ATP bound}
 \Phi_{\rm ATP}^{\rm n} \geq E_{\rm n}\, V +H_{\rm n},\quad  \Phi_{\rm ATP}^{\rm a} \geq E_{\rm a}\, V + H_{\rm a}.
\end{equation}
It follows from the choice of $\bv_1$, $\bv_2$ and (\ref{circle}) that
\[
{\bf j}^{\rm n} =
\left[\begin{array}{c}
j_{1}\\
j_{2}\\
j_{3}
\end{array}\right]=\left[\begin{array}{c}
\alpha_{1}\\
-2\alpha_{1}\\
0
\end{array}\right]+\left[\begin{array}{c}
0\\
1/3\alpha_{2}\\
\alpha_{2}
\end{array}\right]+\frac{1}{2}\left[\begin{array}{c}
J_{1}\\
J_{2}\\
J_{3}
\end{array}\right],
\]
and
\[
{\bf j}^{\rm a} = \left[\begin{array}{c}
j_{7}\\
j_{8}\\
j_{9}
\end{array}\right]=\left[\begin{array}{c}
-\alpha_{1}\\
2\alpha_{1}\\
0
\end{array}\right]+\left[\begin{array}{c}
0\\
-1/3\alpha_{2}\\
-\alpha_{2}
\end{array}\right]+\frac{1}{2}\left[\begin{array}{c}
J_{1}\\
J_{2}\\
J_{3}
\end{array}\right].
\]
Substituting the expression for the components of $\bj^n$ and $\bj^a$ into (\ref{ATP_n})--(\ref{ATP_a}) and  further into (\ref{ATP bound}), we obtain the following conditions for $\alpha_1$ and $\alpha_2$:
\[
 2\,\alpha_1 + \frac {16}3\, \alpha_2 + J_1  +\frac 83\, J_3 \geq E_{\rm n} V + H_{\rm n},
\]
and
\[
 -2\,\alpha_1 - \frac{16}3\, \alpha_2 + J_1  +\frac 83\, J_3 \geq E_{\rm a} V + H_{\rm a},
\]
from which it readily follows that $\alpha_1$ and $\alpha_2$ must satisfy
\begin{equation}\label{feasible region}
 -J_1 - \frac 83\, J_3 + E_{\rm n} V +H_{\rm n} \leq 2\,\alpha_1 + \frac {16}3\, \alpha_2 \leq J_1 + \frac 83 \,J_3 -E_{\rm a} V - H_{\rm a}.
\end{equation}

It follows from the metabolic interpretation of this chain of inequalities that the maximal rate of neuronal activity supported by the model is the value of $V=V^{*}$, at which the upper and lower bounds in (\ref{feasible region}) coincide. A simple manipulation leads to the formula
\begin{equation}\label{Vstar}
 V^* = \frac 2{E_{\rm n} + E_{\rm a}}\left(J_1 + \frac {8}{3} J_3\right) - \frac{H_{\rm n}+H_{\rm a}}{E_{\rm n}+E_{\rm a}}
  = \frac 2{E_{\rm tot}}\left(1 + \frac {8}{3} {\rm OGI}\right){\rm CMR}_{\rm Glc} - \frac{H_{\rm tot}}{E_{\rm tot}},
\end{equation}
where $E_{\rm tot} = E_{\rm n} + E_{\rm a}$ and $H_{\rm tot} = H_{\rm n} + H_{\rm a}$.  Since formula (\ref{Vstar}) assumes that all energy produced is used for signaling and household chores, leaving no room for other energetic expenses, it provides an upper bound for the V-cycle activity, that is,  $V\leq V^*$.

Moreover, starting from equation (\ref{Vstar}) , we can define a theoretical lower bound for the total glucose oxidation as a function of the V-cycle flux. Let ${\rm CMR}_{\rm Glc(ox)}$ denote the flux of glucose oxidized in the unit, measured as one half of the total TCA cycle activity in the system. A simple manipulation shows that
\[
  {\rm CMR}_{\rm Glc(ox)} = \frac 12(\varphi_3 + \varphi_9) =  \frac{\rm OGI}{6}{\rm CMR}_{\rm Glc},
\]
therefore ${\rm CMR}_{\rm Glc(ox)}$ can be expressed in terms of $V^*$ as
\begin{equation}\label{linear relation}
 {\rm CMR}_{\rm Glc(ox)} = \gamma E_{\rm tot}  V^* + \gamma H_{\rm tot}, \quad \gamma =  \frac{{\rm OGI}}{12+ 32\,{\rm OGI}}.
\end{equation}
This formula expresses the experimentially established linear dependence of the brain's energy demand on the V-cycle activity, see: \cite{Sibson1998} for rodent data, and  \cite{Gruetter2001,Lebon2002,Shen1999} for human data. Further discussion on this topic can be found in \cite{Gjedde2002,Hyder2006} and in \cite{Calvetti2012}, where the dependency was addressed in the light of model simulations.

\subsubsection{Estimates from the Literature: Rodent Data}

To shed some light on constituents of this formula, we discuss the ``bottom up'' approach to cerebral energetic needs first proposed in \cite{Attwell2001}, and later updated in \cite{Howarth2012}. In \cite{Attwell2001}, the energy need of the brain is estimated by analyzing the basic processes constituting the brain signaling and maintenance. The estimated aggregate amount of ATP molecules required for each glutamate molecule passing through the V-cycle is of the order of 41 ATP/glutamate distributed among neurotransmitter recycling ($\sim 3$ ATP), postsynaptic processing ($\sim 35$ ATP), and presynaptic processing ($\sim 3$ ATP). The cost of action potential propagation initially estimated to be around 48 ATP/glutamate, was later downgraded  in \cite{Howarth2012} to about one third,  or 16 ATP/glutamate. These estimates yield a total stoichiometric estimate of $E_{\rm tot} = 57$. Assuming an OGI of the order of 5--6, the coefficient $\gamma$ ranges in the interval 0.0291--0.0294, leading to a coefficient value $\gamma E_{\rm tot} \sim$ 1.66--1.68.

The household energy demand of a cortical neuron, according to \cite{Howarth2012}, is approximately 20--25\% of the total energy at a low activity level, corresponding to a firing frequency of $4\;{\rm Hz}$. Denoting the corresponding V-cycle activity by $V_0$, we introduce the energy partitioning index (EPI), defined as
\begin{equation}\label{EPI1}
{\rm EPI} = \frac{\mbox{glucose oxidized for household maintenance}}{\mbox{total glucose oxidized at $V=V_0$}},
\end{equation}
which, in the light of (\ref{linear relation}), yields
\begin{equation}\label{EPI2}
{\rm EPI} = \frac{H_{\rm tot}}{E_{\rm tot} V_0 + H_{\rm tot}},
\end{equation}
and further, by solving for $H_{\rm tot}$, we obtain the estimate
\begin{equation}\label{Htot}
 H_{\rm tot} = \underbrace{\frac{{\rm EPI}}{1 - {\rm EPI}}}_{=\beta} E_{\rm tot} V_0 = \beta E_{\rm tot}V_0.
\end{equation}
Hence, by using an estimate for $V_0$ that corresponds to awake resting state of the human brain, $V_0 = 0.25\,\mu\,{\rm mol}/{\rm min}$ per one gram tissue, and the aforementioned estimate for ${\rm EPI}\sim$1/5--1/4, the estimate for household energy (measured in terms of ATP hydrolysis flux) is of the order of $H_{\rm tot}\sim 3.56$--$4.75\;\mu\,{\rm mol}/{\rm min}$ per one gram tissue, and the corresponding glucose oxidation flux needed to maintain the household functions is  $\gamma H_{\rm tot} \sim0.10$ -- $0.14\;\mu\,{\rm mol}/{\rm min}$ per one gram tissue.

The energy estimation in \cite{Attwell2001,Howarth2012} is based mostly on analysis of glutamatergic neuronal activity in rodent brain, although the authors suggest corrections for human brain.

\subsubsection{Inverse Calculations with Human Data}

\begin{table}
\caption{\label{tab:input} Input values used in the inverse problem and the corresponding sources. $^1$Lebon et al., \cite{Lebon2002}.  $^2$This is an estimate; the value refers to the activity level in Howarth et al. and Attwell et al. \cite{Attwell2001,Howarth2012}  $^3$This is implicitly deduced from \cite{Howarth2012}. The OGI value is a typical value reported in literature.}
\begin{center}
\begin{tabular}{lc}
\hline
OGI & 5.4 \\
V-cycle [$\mu\,{\rm mol}/{\rm min}$] & $0.32\pm 0.07$  \\
${\rm CMR}_{\rm Glc(ox)}$ neuron$^1$ & $0.8/2 = 0.4$ \\
${\rm CMR}_{\rm Glc(ox)}$ astrocyte$^1$ & $0.14/2 = 0.07$ \\
Low activity V-cycle$^2$ $V_0$ [$\mu\,{\rm mol}/{\rm min}$]  & $0.25$ \\
${\rm EPI}^3$ & 0.2\\
\hline
\end{tabular}
\end{center}
\end{table}

From the point of view of the model, it is of interest to consider the following inverse problem:  {\em Estimate the cell-level energetic parameters from measured data, using the relation between  ${\rm CMR}_{\rm Glc(ox)}$ and the rate of the V-cycle.} In particular, we focus here on measurements in human brain, reported in \cite{Lebon2002},  where the estimated TCA cycle fluxes per one gram tissue are given as $0.80\;\mu\,{\rm mol}/{\rm min}$ in neuron and $0.14\;\mu\,{\rm mol}/{\rm min}$ in glia, respectively, at the V-cycle activity level $V = 0.32\pm 0.07\;\mu\,{\rm mol}/{\rm min}$.
Bearing in mind that ${\rm CMR}_{\rm Glc(ox)}$ is one half of the total TCA cycle activity, we obtain the values given in Table~\ref{tab:input}.
Using these values, the ${\rm CMR}_{\rm Glc}$ is
\[
 {\rm CMR}_{\rm Glc} = \frac{6}{\rm OGI}  {\rm CMR}_{\rm Glc(ox)}  = 0.5222\,\mu{\rm mol}/{\rm min} \mbox{ per 1 gram tissue.}
\]
By substituting (\ref{Htot}) into (\ref{linear relation}), we obtain
\[
 {\rm CMR}_{\rm Glc(ox)} = \gamma E_{\rm tot}  ( V^* + \beta V_0),
\]
which allows us to solve for $E_{\rm tot}$,
\[
 E_{\rm tot} = \frac{ {\rm CMR}_{\rm Glc(ox)} }{ \gamma   ( V^* + \beta V_0)}.
\]
To fix the value $V^*$, the maximum possible V-cycle rate allowed by this uptake, we use the upper bound reported in \cite{Lebon2002}, listed in Table~\ref{tab:input},
\[
 V^* = 0.32 + 0.07 = 0.39,
\]
which gives the relative V-cycle activity index (RVAI),
\[
 {\rm RVAI} = \frac{V}{V^*} = 0.82.
\]
With this value, we find the total energy demand, rounding off to nearest integer, to be
\[
 E_{\rm tot} = 36
\]
which, upon substitution into formula (\ref{Htot}), yields the approximation
\[
 H_{\rm tot}  = 2.25 \,\mu\,{\rm mol}/{\rm min} .
\]
In the following, we assume that roughly one half of the household energy is used in glia, one half in neuron, leading to
\[
 H_{\rm a} = H_{\rm n} \approx 1.13 \,\mu\,{\rm mol}/{\rm min}.
\]
We collect the computed estimates in Table~\ref{tab:derived}.

\begin{table}
\caption{\label{tab:derived} Derived quantities based on the model and estimated values given in Table~\ref{tab:input}.}
\begin{center}
\begin{tabular}{lc}
\hline
${\rm RVAI}$ & 0.82 \\
$E_{\rm tot}$ & 36 \\
$E_{\rm a}$ & 5 \\
$E_{\rm n}$ & 31 \\
$H_{\rm tot}$ $[mu\,{\rm mol}/{\rm min}]$ &  2.25 \\
$H_{\rm a}$ $[mu\,{\rm mol}/{\rm min}]$ &  1.13 \\
$H_{\rm n}$ $[mu\,{\rm mol}/{\rm min}]$ &  1.13 \\
\hline
\end{tabular}
\end{center}
\end{table}

\subsection{Glucose Partitioning and Lactate Flux}

It is of interest to interpret the coefficients  $\alpha_1$ and $\alpha_2$ in terms of glucose, lactate and oxygen uptake of each cell type. To shed some light on this, we recall that the first components of $\bj^{\rm n}$ and $\bj^{\rm a}$ are the rate of glucose uptake in the corresponding cells, and
\[
 j_1  =  \alpha_1 + \frac 12 J_1,\quad
 j_7  = -\alpha_1 + \frac 12 J_1,
\]
 $J_1$ being the total glucose uptake. When $\alpha_1 = 0$, the glucose is evenly partitioned between neuron and astrocyte; a positive $\alpha_1$ shifts the glucose uptake more towards neuron, and a negative $\alpha_1$ towards astrocyte.  Similarly, recalling that the third components of $\bj^{\rm n}$ and $\bj^{\rm a}$ correspond to the oxygen uptake of the corresponding cell, and that
\[
 j_3  = \alpha_2 + \frac 12 J_3,\quad
 j_9  = -\alpha_2 + \frac 12 J_3,
\]
it follows that a large value of $\alpha_2$ corresponds to highly oxidative neuron, while a large negative $\alpha_2$ shifts the oxidative activity more towards the astrocyte. Summarizing, glucose and oxygen partitionings are regulated by $\alpha_1$ and $\alpha_2$, respectively. The lactate partitioning, on the other hand, is a function of the glucose and oxygen partitioning, since
\[
 j_2 = -2\,\alpha_1 + \frac 13 \alpha_2 + \frac 12 J_2,\quad
 j_8 =  2\,\alpha_1 - \frac 1 3 \alpha_2 + \frac 12 J_2.
\]
Note that when ${\rm OGI}<6$, the neuron-astrocyte complex is a lactate producer, because $J_2<0$, meaning that lactate is being released. The neuron takes up lactate when $j_2>0$, which, in turn, implies that
\begin{equation}\label{ANLS}
 -2\,\alpha_1 + \frac 13 \alpha_2 > - \frac 12 J_2.
\end{equation}
Observe that when this happens, with ${\rm OGI}<6$, we have $j_8<0$, implying that the astrocyte produces and releases lactate. We refer to this situation as having the system in the ANLS (astrocyte-neuron lactate shuttle)  state. Conversely, if $j_8>0$, or
\begin{equation}\label{NALS}
2\,\alpha_1 - \frac 13 \alpha_2 > - \frac 12 J_2,
\end{equation}
the astrocyte takes up the lactate produced by the neuron: in this case we say that the system is in NALS (neuron-astrocyte lactate shuttle) state. It is possible that both $j_2$ and $j_8$ are negative, in which case both cells produce lactate.

\begin{figure*}
\centerline{
\includegraphics[width=4cm]{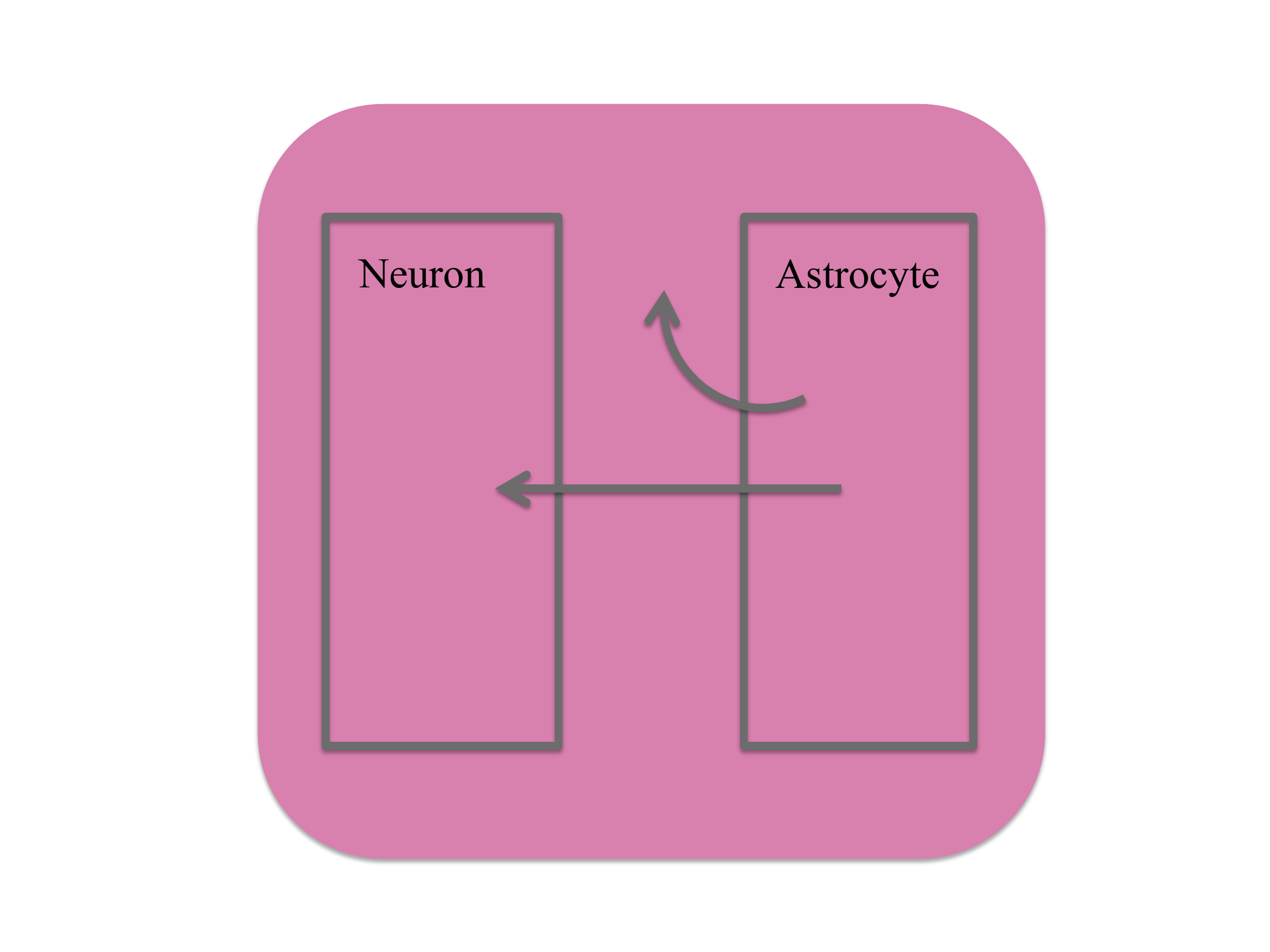}
\includegraphics[width=4cm]{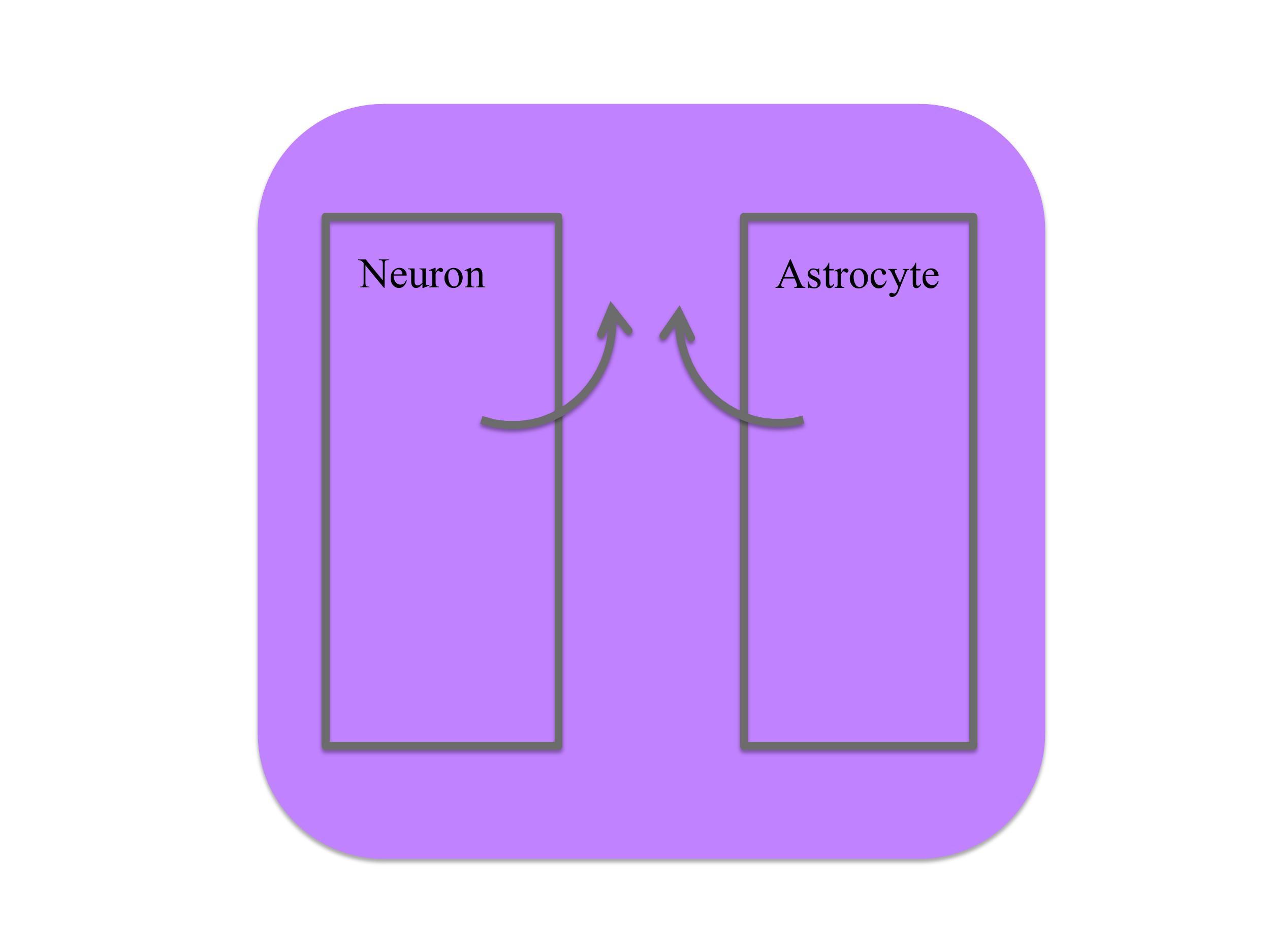}
\includegraphics[width=4cm]{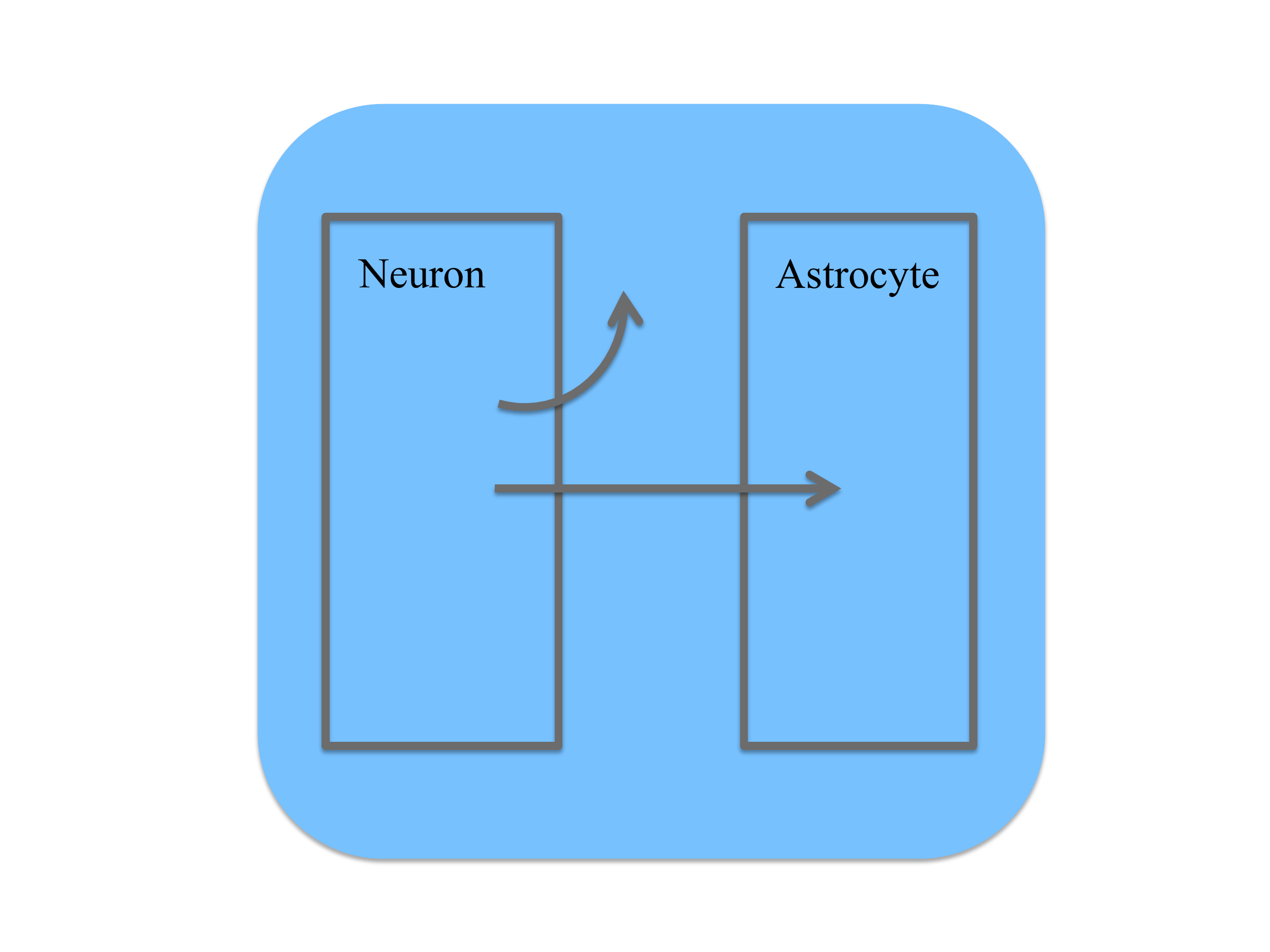}}
\centerline{
\includegraphics[width=4cm]{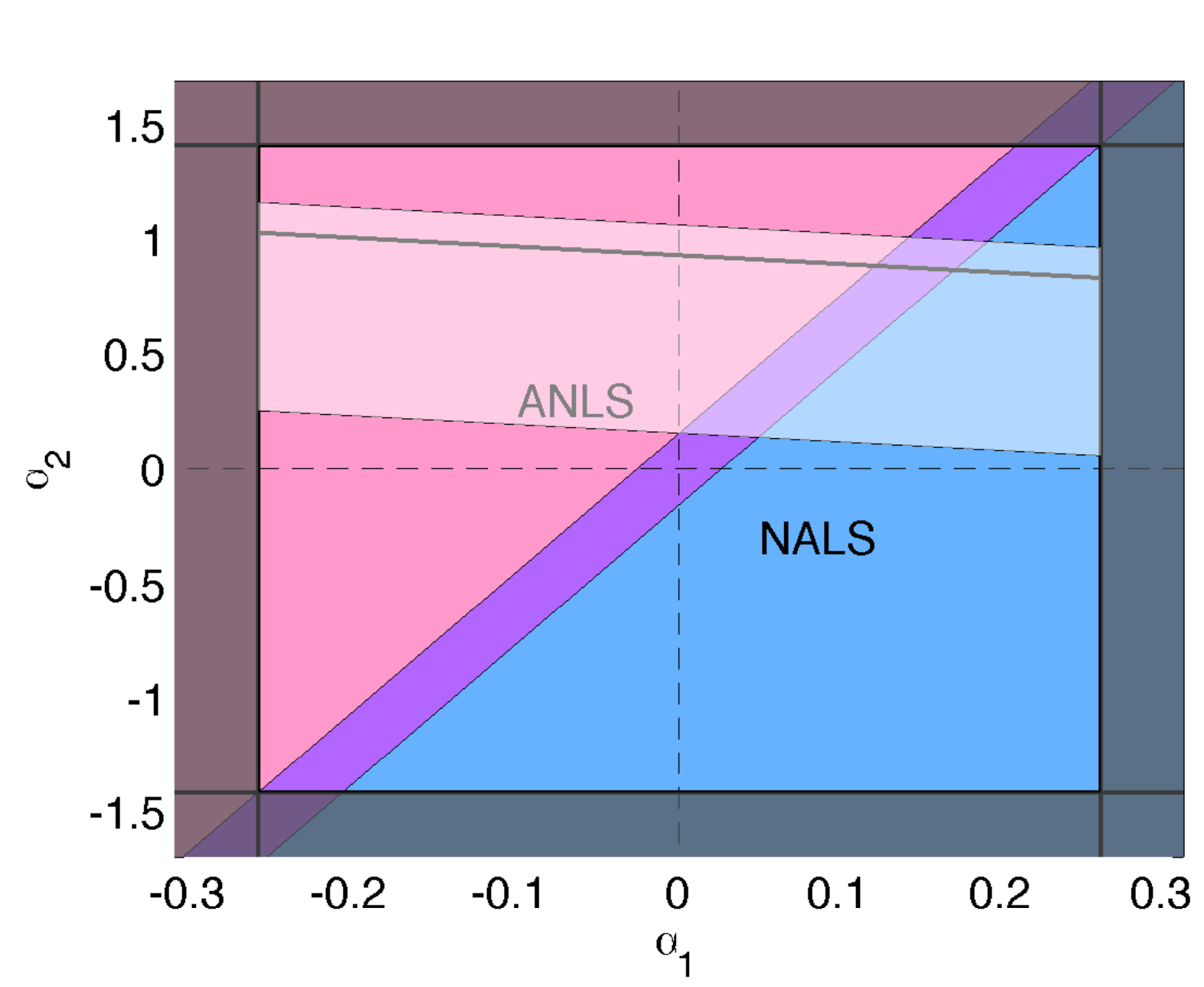}
\includegraphics[width=4cm]{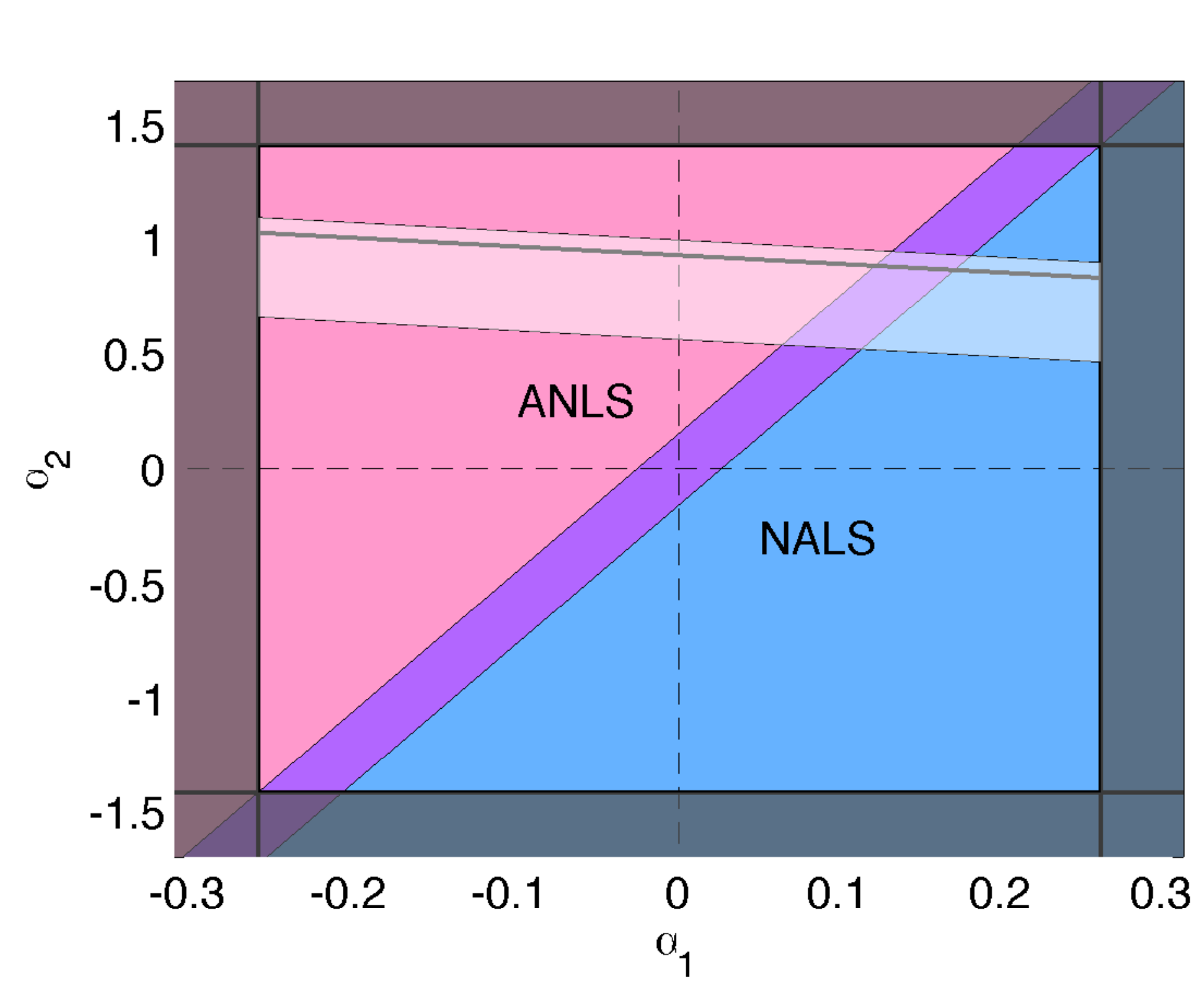}
\includegraphics[width=4cm]{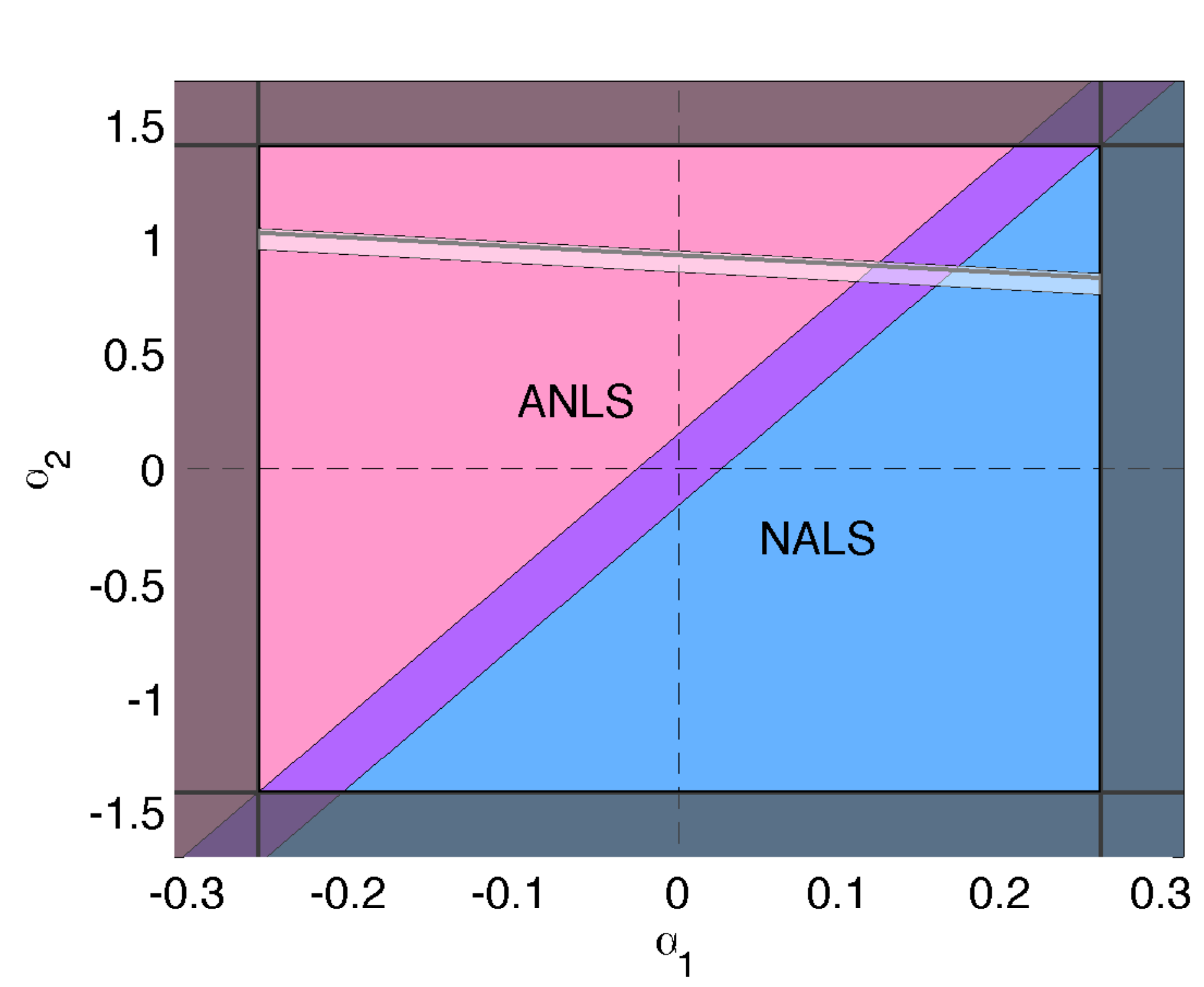}}
\centerline{
\includegraphics[width=4cm]{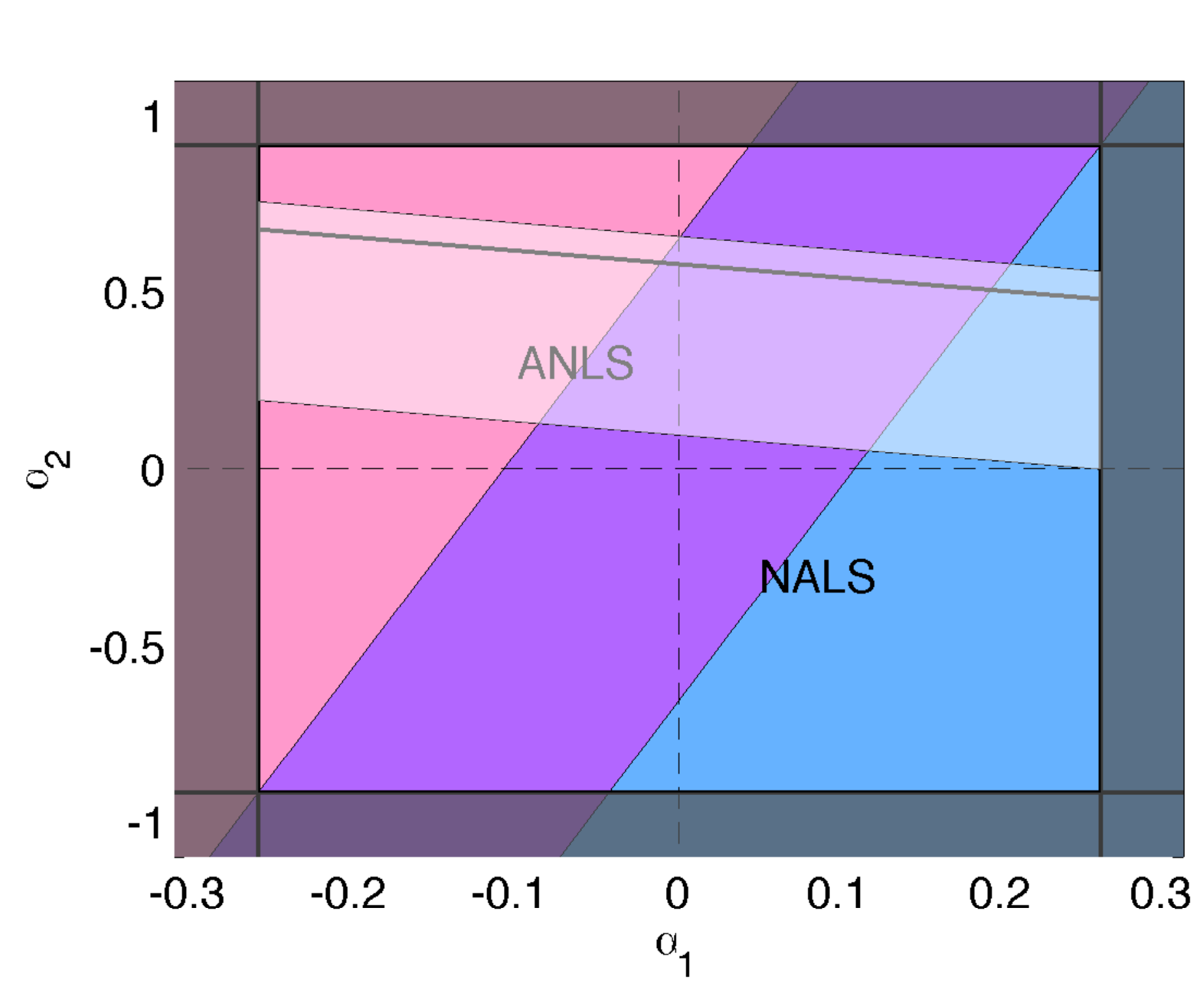}
\includegraphics[width=4cm]{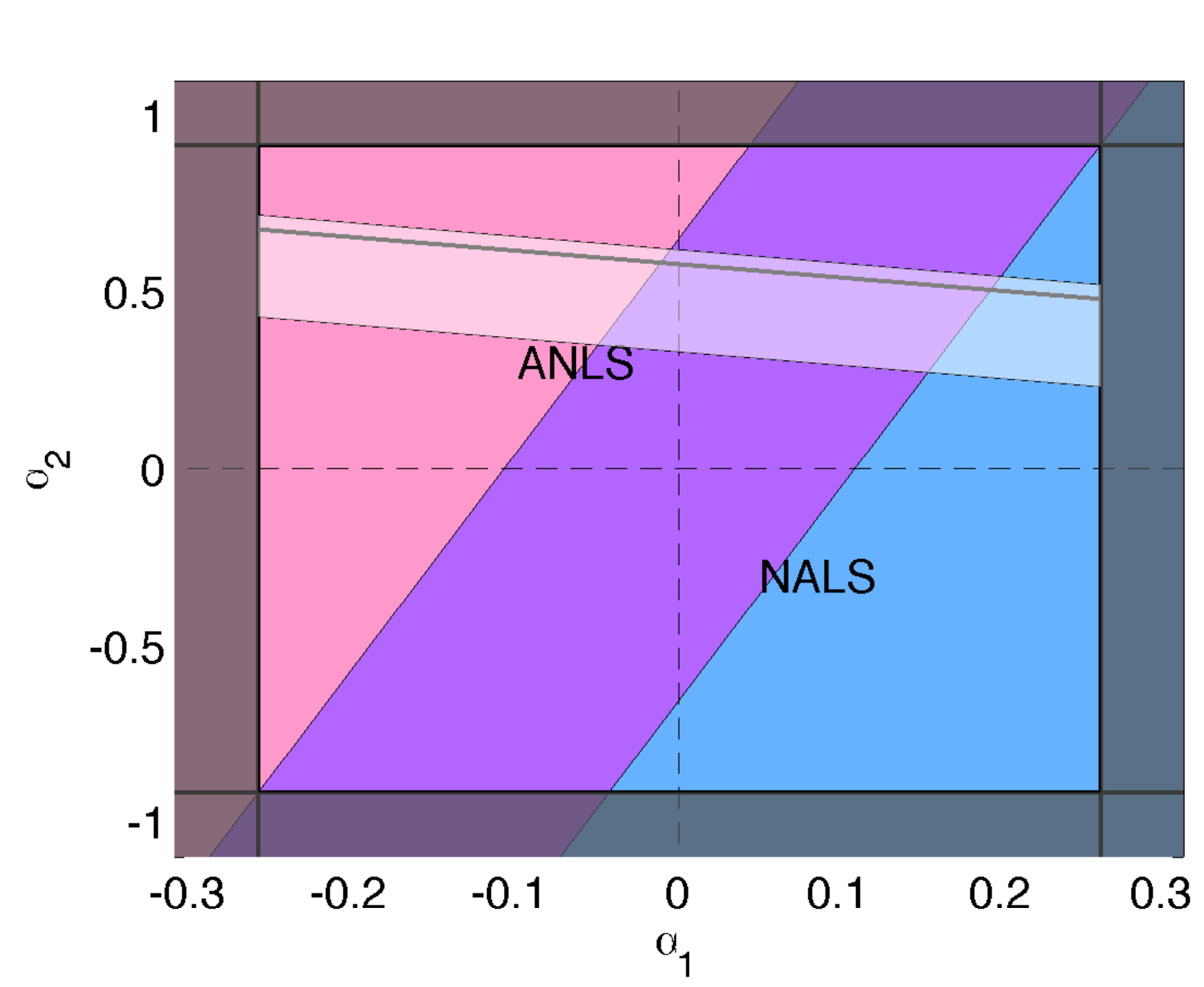}
\includegraphics[width=4cm]{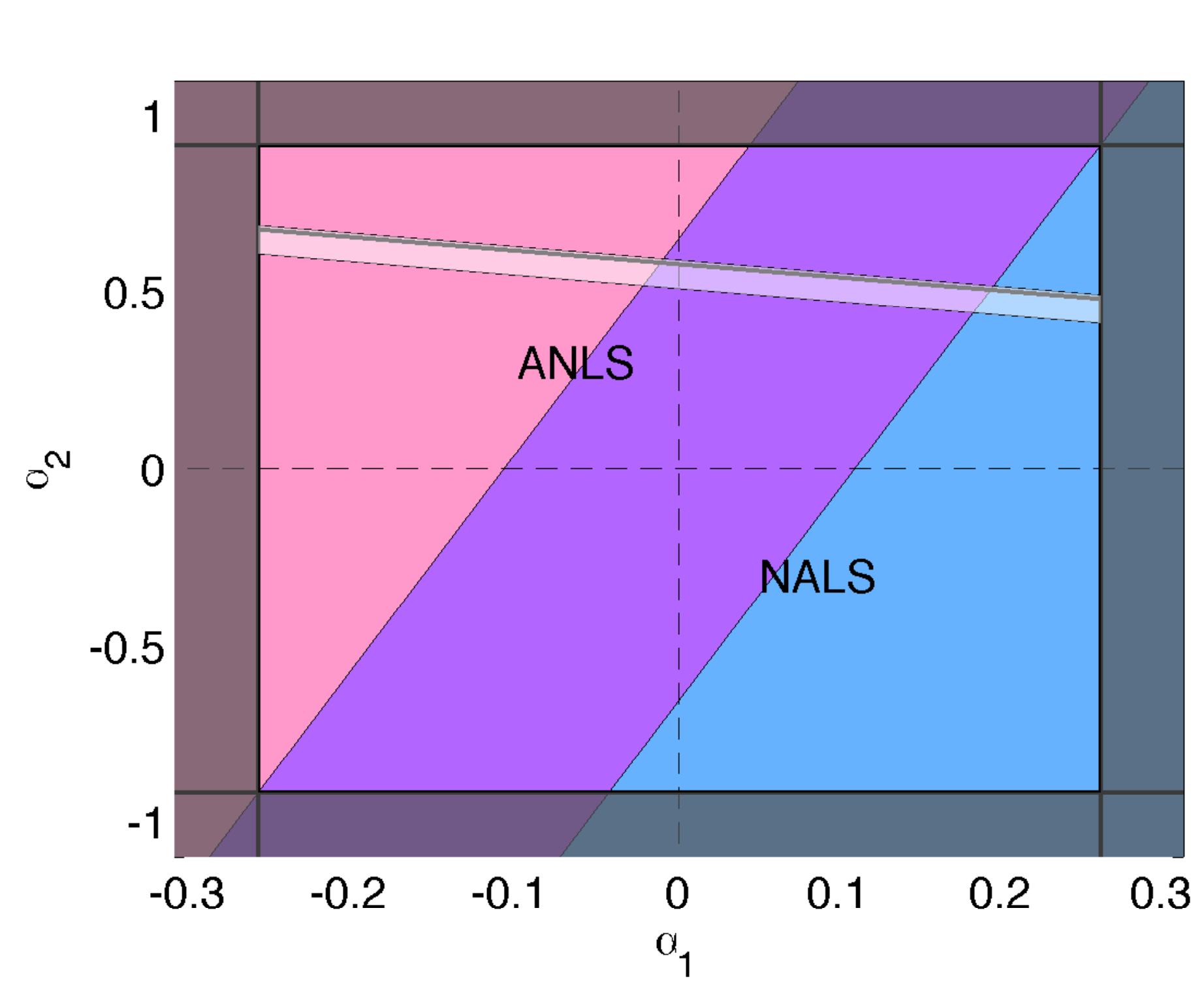}}
\caption{\label{fig:feasible}  {\bf Top row:} Schematics of the three possible lactate traffic configurations when ${\rm OGI}<6$. On the left, astrocyte produces lactate, while neuron takes it up (ANLS), in the center, both cells produce lactate, and on the right, neuron is the lactate producer and astrocyte oxidizes it (NALS).
{\bf Center row:}  The feasible regions for the parameters $\alpha_1$ and $\alpha_2$ with three different activity levels. In the plots, the parameter values defining the metabolite uptakes are  given in Tables~\ref{tab:input}-\ref{tab:derived}, with varying V-cycle activity as follows: $V = V_0 = 0.25\mu\,{\rm mol}/{\rm min}$ (left), $V = 0.32\; \mu\,{\rm mol}/{\rm min}$ (center), and $V = 0.95\,V^* = 0.37\; \mu\,{\rm mol}/{\rm min}$ (right).
In the plots,  the coloring of different regions is in agreement with that of the different configurations  in the top row. The energetically feasible region in which the inequalities (\ref{feasible region}) are valid is highlighted. The black line in the feasible region corresponds to $V=V_*$, i.e., the energy production and use are tightly coupled. The values $\alpha_1=0$ and $\alpha_2=0$ indicated with dashed lines  correspond to equal partitioning of glucose and oxygen, respectively. Observe that equal glucose partitioning is energetically feasible almost exclusively in the ANLS region.
{\bf Bottom Row:} As in the middle row, but the value ${\rm OGI = 3.5}$ is used. From left to right, we use the same RVAI values ($0.64$ ,$0.82$, $0.95$, respectively) as in the second row. The maximum feasible V-cycle value in this case is $V^*=0.24 \; \mu\,{\rm mol}/{\rm min}$.
}
\end{figure*}

Figure~\ref{fig:feasible} shows how the different choices of $\alpha_1$ and $\alpha_2$ determine whether the system is in NALS, ANLS, or in an in-between state, color coding the different regions of the rectangle representing all possible choices of $(\alpha_1,\alpha_2)$. The energetically feasible area, highlighted in the figure, corresponds to values of that satisfy the inequality constraints (\ref{feasible region}), while the values of $(\alpha_1,\alpha_2)$ for which $V=V^*$ are on the black almost horizontal line across this area. Observe that when $V<V^*$ we implicitly assume that some of the energy produced by the cells is used for unspecified activities others than signaling and household. Alternatively, this can be interpreted as a way to introduce uncertainty in the household energy level. The larger the uncertainty, the wider the feasible region becomes.  A particular feature to be pointed out is that, in particular when $V$ is close to or equal to $V^*$, hence the feasible region becomes narrow, the direction of the lactate traffic between the cells can be readily deduced from the glucose partitioning, or $\alpha_1$. This observation
is in line with the findings in \cite{Calvetti2012,Massucci2013}.

\section{Multiple Units}

\begin{figure*}
\centerline{
\includegraphics[width=10cm]{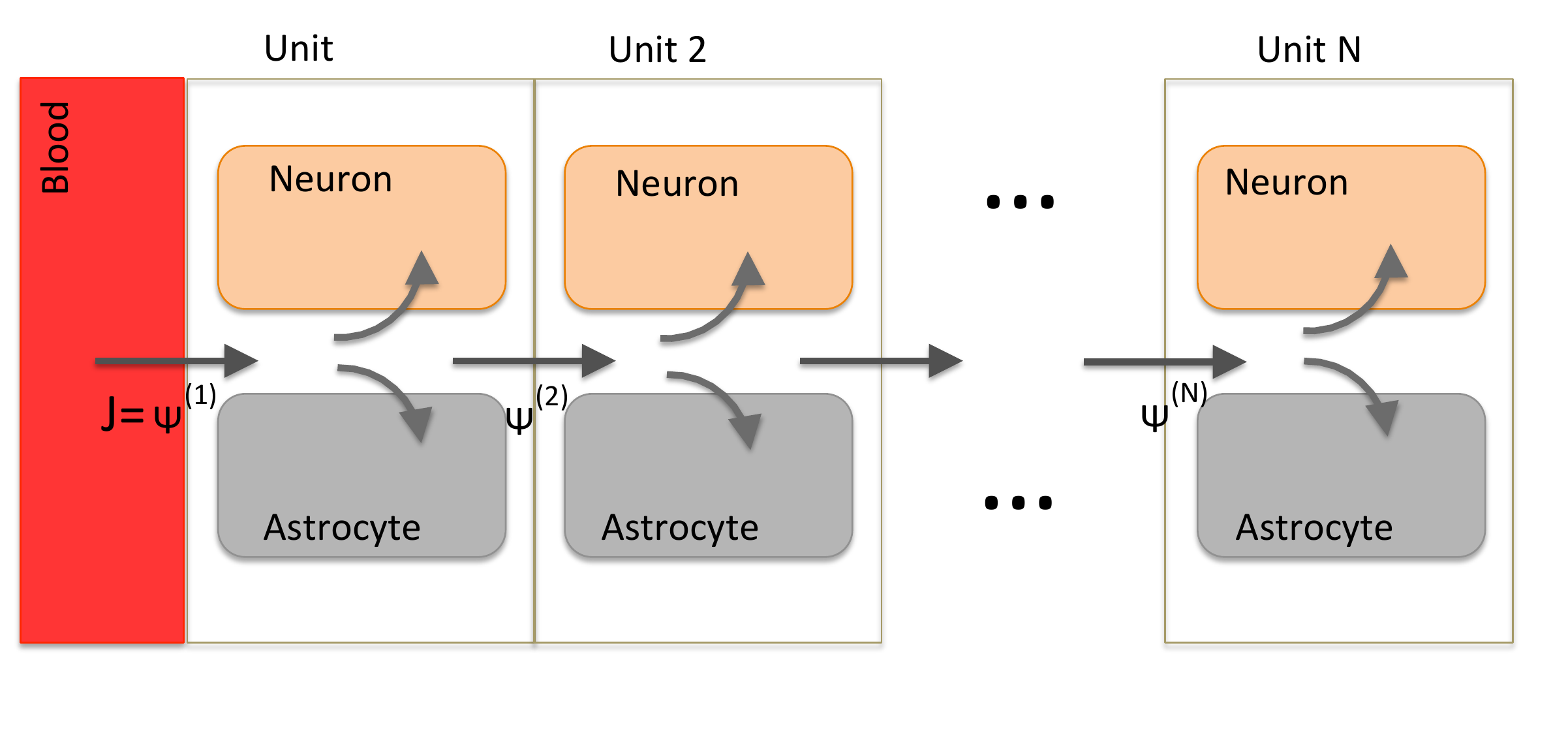}
}
\caption{\label{fig:unit configuration} A schematic picture of the subunit configuration. The arrows indicate the diffusion in the ECS of the tissue, as well as the metabolite uptake by the cell compartments in the units.}
\end{figure*}

The model in the previous section is analytically tractable because of the simplified biochemistry and geometry. When working with more realistic models it becomes necessary to employ sophisticated computational techniques, e.g., carrying out a Bayesian flux balance analysis (BFBA) to shed light on possible steady state configurations \cite{Heino,Metabolica}. This approach has been used by the authors to study complex biochemical models for brain metabolism; see, e.g., \cite{Occhipinti2010,Calvetti2012,Calvetti2013}; In this work, we extend the analysis towards a geometrically distributed  model. More precisely, the model consists of $N$ identical neuron/astrocyte complexes, communicating with each other through a common extracellular space. We assume that the units are aligned in a one-dimensional array, with the unit $n=1$ closest to a capillary, the unit $n=N$ most distant from it. A schematic picture of the geometric setting is shown in Figure~\ref{fig:unit configuration}.  The model can be regarded as a mathematical idealization of a more realistic spatially distributed geometry, namely a Krogh cylinder around a single capillary, and can be derived through a model reduction process described in \cite{Calvetti2014}. In this work, the model serves as a proof of concept to elucidate the complexities introduced to energy metabolism analysis when moving beyond the spatial lumping paradigm.

\subsection{Subunit Stoichiometry}

To set up the model, consider a single $k$th unit, $1\leq k\leq N$. As in the lumped model, the unit is characterized by the 12 reactions indicated in Figure~\ref{fig:small} and listed in Table~\ref{tab:reactions}, and 12 transfers from ECS into the cells, given in Table~\ref{tab:transports}. We collect the transport and reaction fluxes in a single vector,
\[
 \bu^{(k)} = \left[\begin{array}{c}  \bvarphi^{(k)} \\ \bj^{(k)}  \end{array}\right]\in\R^{24},\quad \mbox{where}\quad \bvarphi^{(k)}\in\R^{12}\quad \bj^{(k)} \in \R^{12}.
\]
Within each unit, the mass balance conditions define 20 stoichiometric linear equations, listed in Tables~\ref{tab:transports}--{\ref{tab:stoichiometry}, thus defining a stoichiometric matrix ${\mathsf A} \in\R^{20\times 24}$ such that
\begin{equation}\label{stoichiometry single}
 {\mathsf A} \bu^{(k)} = 0,\quad1\leq k\leq N.
\end{equation}
Since each subunit complex is independent, we express the reaction portion of the mass balance equations for all subunits in the aggregate matrix equations,
\begin{equation}\label{inside units}
 \big({{\mathsf I}}_N\otimes {\mathsf A}\big){\bf u} = 0, \quad {\bf u} = \left[\begin{array}{c} {\bf u}^{(1)} \\ \vdots \\ {\bf u}^{(N)}\end{array}\right] \in\R^{22 N},
\end{equation}
where ${\mathsf I}_N$ is the $N\times N$ unit matrix and ``$\otimes$'' stands for the Kronecker product of the matrices, so that ${{\mathsf I}}_N\otimes {\mathsf A} \in\R^{20 N\times 24 N}$.

\subsection{Coupling Subunits by Diffusion}

Consider next the diffusion. Our model assumes that adjacent units interchange through a diffusion process the four metabolites:  glucose, lactate, oxygen, and carbon dioxide.
For $k=1,\cdots,N$, we define a diffusion flux vector,
\[
\bpsi^{(k)}=\left[\begin{array}{c}
\psi_{1}^{(k)}\\
\psi_{2}^{(k)}\\
\psi_{3}^{(k)}\\
\psi_{4}^{(k)}
\end{array}\right],
\]
 the vector containing the diffusion fluxes of glucose, lactate, oxygen and carbon dioxide in this order from the ECS of the $(k-1)$th subunit to the ECS of the  $k$th subunit, with the convention that $k=0$ corresponds to the diffusion fluxes from the blood vessel to the first ECS subunit, and $\psi_i^{(k)}>0$ means net flux towards the $k$th unit, see Figure~\ref{fig:unit configuration}.

The net flux of metabolites into the $k$th compartment is given by
\[
 \bpsi^{(k)}_{\rm net} = \bpsi^{(k)} - \bpsi^{(k+1)},\quad \bpsi^{(N+1)} = 0,
\]
and to maintain the mass balance in the ECS of each subunit, this net flux must coincide with the uptake/production by the neurons and astrocytes, i.e.,
\begin{equation}\label{diffusion balance}
 \bpsi^{(k)}_{\rm net} = \left[\begin{array}{c} j^{(k)}_1 \\ j^{(k)}_2\\j^{(k)}_3\\j^{(k)}_4\end{array}\right] + \left[\begin{array}{c} j^{(k)}_7 \\ j^{(k)}_8\\j^{(k)}_9\\j^{(k)}_{10}\end{array}\right] = {\mathsf B} \bu^{(k)},
\end{equation}
where ${\mathsf B} \in\R^{4\times 24}$ is an appropriately defined matrix effectuating the above stoichiometry. To organize these equations in a single matrix equation, define
\[
 \bpsi = \left[\begin{array}{c} \bpsi^{(1)} \\ \vdots \\ \bpsi^{(N)}\end{array}\right] \in\R^{4N},
\]
and further,
\[
{\mathsf P}=\left[\begin{array}{rrrr}
1 & -1\\
 & 1 & \ddots\\
 &  & \ddots & -1\\
 &  &  & 1
\end{array}\right]\in \R^{N\times N},
\]
so that the system of equations (\ref{diffusion balance}) can be arranged into a single matrix equation,
\begin{equation}\label{diffusion}
\big( {\mathsf P}\otimes{\mathsf I}_4\big)\bpsi = \big({\mathsf I}_N\otimes{\mathsf B}\big)\bu.
\end{equation}
Finally, we need to couple the stoichiometric equations to the CMR of the entire unit. We define the net flux vector of glucose, lactate, oxygen, and carbon dioxide from blood to tissue,
\[
\bJ = \left[\begin{array}{c} J_1 \\ J_2 \\ J_3 \\ J_4\end{array}\right],
\]
with the sign convention that positive sign indicates flux from blood to tissue. We set the boundary condition in matrix form as
\begin{equation}\label{CMR}
 \bJ = \bpsi^{(1)} =   \big(\be_1\otimes {\mathsf I}_4\big)\bpsi,\quad \be_1 = \left[\begin{array}{cccc} 1 & 0 & \cdots & 0\end{array}\right]\in\R^{1\times N}.
\end{equation}

We are now ready to assemble the stoichiometry into a single matrix equation. By defining
\[
 \bX = \left[\begin{array}{c} \bpsi \\ \bu \end{array}\right] \in\R^{4N + 24 N},
\]
the stoichiometry is given by the linear system combining equations (\ref{inside units}), (\ref{diffusion}) and (\ref{CMR}) as
\begin{equation}\label{combined stoichiometry}
{\mathsf M}\bf X =
 \left[\begin{array}{cc} \be_1\otimes {\mathsf I}_4 & 0 \\
  0 & {\mathsf I}_N\otimes{\mathsf A} \\ {\mathsf P}\otimes {\mathsf I}_4 & - {\mathsf I}_N\otimes{\mathsf B} \end{array}\right] \bX = \left[\begin{array}{c} \bJ \\ 0\end{array}\right] \in \R^{4 + 20 N + 4N}.
\end{equation}
The properties of the system matrix ${\mathsf M}\in\R^{(24N+4)\times 28N}$ are analyzed below.

\subsection{Bound Constraints}

In addition to the stoichiometry (\ref{combined stoichiometry}), the vector $\bX$ has to comply with the a priori bound constraints. In each unit, we have the bounds for the reaction fluxes,
\[
 \left[\begin{array}{c} \varphi_1^{(k)} \\ \varphi_3^{(k)} \\ \varphi_4^{(k)} \\ \varphi_5^{(k)}
 \end{array}\right] \geq 0,
 \quad
 \left[\begin{array}{c} \varphi_7^{(k)} \\ \varphi_9^{(k)} \\ \varphi_{10}^{(k)} \\ \varphi_{11}^{(k)}
  \end{array}\right] \geq 0,
  \quad
  \varphi_6^{(k)} \geq H_{\rm n}^{(k)},
  \quad
  \varphi_{12}^{(k)}\geq H_{\rm a}^{(k)},
\]
where $H_{\rm n}^{(k)}$ and $H_{\rm a}^{(k)}$ are the household energies of the compartments in the $k$th unit.
Similarly, for transport fluxes we have the bound constraints
\[
 \left[\begin{array}{r} j_1^{(k)} \\ j_3^{(k)} \\ -j_4^{(k)} \\ -j_5^{(k)} \\ j_6^{(k)}\end{array}\right] \geq 0,\quad  \left[\begin{array}{r} j_7^{(k)} \\ j_9^{(k)} \\ -j_{10}^{(k)} \\ j_{11}^{(k)} \\ -j_{12}^{(k)}\end{array}\right] \geq 0.
\]
We express all the conditions into a single matrix inequality,
\begin{equation}\label{bound constraints}
 {\mathsf C}\bX \geq \bc, \quad {\mathsf C} \in \R^{20N\times (4N+24N)},
\end{equation}
which is to be understood component-wise.

\subsection{Degrees of Freedom}
To analyze the degrees of freedom of the distributed model, we consider the system matrix ${\mathsf M}$ given in (\ref{combined stoichiometry}). The matrix ${\mathsf A}\in\R^{20\times 24}$ has ${\rm Rank}({\mathsf A}) = 19$ implying that the null space ${\mathcal N}(\mathsf A)$ has dimension $24-19=5$. In the light of the single unit analysis, two of them can be associated to the glucose and lactate partitioning between the neuron and the astrocyte, one to the unspecified V-cycle activity, and two for the underdetermined ATP demand in each cell. Introduce the notation
\[
 {\mathcal N}({\mathsf A}) = {\rm span}\big\{\bv_1,\bv_2,\bv_3,\bv_4,\bv_5\big\},\quad \bv_j\in\R^{24}.
\]
If $\bX$ is a solution of the system (\ref{combined stoichiometry}), then it must be of the form
\[
 \bX = \left[\begin{array}{c} \bpsi \\ \bu \end{array}\right],\quad \bu = \sum_{k=1}^N \sum_{j=1}^5 \alpha^{(k)}_j \be_k\otimes \bv_j,
\]
where $\be_k\in\R^N$ is the canonical unit basis vector. This follows from the fact that each component $\bu^{(k)}$ in (\ref{inside units}) needs to satisfy the condition ({\ref{stoichiometry single}). It therefore follows that
\[
 \big({\mathsf P}\otimes {\mathsf I}_4\big)\bpsi = \big({\mathsf I}_N\otimes {\mathsf B}) \bu = \left[ \begin{array}{c} \sum_{j=1}^5\alpha^{(1)}_j {\mathsf B}\bv_j \\ \vdots \\
 \sum_{j=1}^5\alpha^{(N)}_j {\mathsf B}\bv_j\end{array}\right],
\]
which allows us to solve recursively for the diffusion fluxes,
\[
 \bpsi^{(\ell)} = \sum_{k=\ell}^N \sum_{j=1}^5\alpha^{(k)}_j {\mathsf B}\bv_j,\quad 1\leq \ell\leq N.
\]
In particular, it follows that
\[
 \bJ = \bpsi^{(1)} =  \sum_{k=1}^N \sum_{j=1}^5\alpha^{(k)}_j {\mathsf B}\bv_j \]
 thus  $ \bJ \in {\mathsf B}\big({\mathcal N}(\mathsf A)\big)$, a subspace of $\R^4$ which can be shown to be of  dimension two. This implies, in particular, that in order for the system to have a solution, the flux $\bJ$ cannot be assigned arbitrarily, but has only two degrees of freedom. As in the case of a single unit, the degrees of freedom can be chosen to be ${\rm CMR}_{\rm Glc}$ and ${\rm OGI}$, leading to the representation
\begin{equation}\label{input}
 \bJ = \left[\begin{array}{c} 1 \\ -2+{\rm OGI}/3  \\ \phantom{-}{\rm OGI} \\  -{\rm OGI}\end{array}\right]{\rm CMR}_{\rm Glc},
\end{equation}
which reflects the stoichiometry of (\ref{blood stoichiometry}).

Based on the considerations above, one would expects the dimension of the null space of ${\mathsf M}$ to be $5N-2$, five degrees of freedom for each unit, minus two corresponding to the two boundary conditions. Numerical test demonstrates that this reasoning is indeed correct.

\subsection{Computational Analysis}

Summarizing the analysis of the previous sections, the traditional flux balance analysis (see \cite{Kauffman2003}) comprises identifying the set of all vectors $\bX\in\R^{28N}$ satisfying
\begin{equation}\label{MX=R}
 {\mathsf M}\bX = \bR = \left[\begin{array}{c}\bJ \\ 0\end{array}\right] \mbox{subject to ${\mathsf C}\bX \geq \bc$,}
 \end{equation}
 where $\bJ$ satisfies the condition (\ref{input}) for some ${\rm CMR}_{\rm Glc}>0$ and ${\rm OGI}>0$. The solution set is a convex polytope, which can be fully described by its vertices. This is the idea behind the Extreme Pathway (ExPa) analysis \cite{Schilling2000,Papin2004}. An alternative way, advocated by the authors (see, e.g., \cite{Heino,Metabolica})  is to explore the solution set by random sampling, by generating an ensemble of possible solution vectors that are used to identify particular properties of the solution set. The statistical sampling method of choice is based on Markov Chain Monte Carlo (MCMC) algorithms, and in particular Gibbs sampling and Hit-and-Run sampling. For examples of different published sampling strategies, see, e.g., \cite{Wiback2004,Massucci2013}.

 The approach, referred to as Bayesian flux balance analysis (BFBA), is based on a stochastic extension of the model: The equality constraint in (\ref{MX=R}) constitutes a basis for a Gaussian likelihood model, that is, $\bR$ is considered as an observation with a probability density conditional on $\bX$,
 \[
  \bR\mid \bX \sim {\mathcal N}(\bX,{\mathsf\Sigma}).
 \]
 Here, ${\mathsf\Sigma}$ is a symmetric positive definite matrix, which is often chosen to be diagonal. The diagonal entries are the variances of each scalar equation constituting the stoichiometric equations. The interpretation of the stochastic extension therefore is that we do not expect the system to obey strictly the stoichiometry, but allow a small variations around the target value. The variance can be interpreted as uncertainties in the model, due, e.g., to stoichiometric simplifications and approximations, slight deviations from the steady state of the system, or uncertainties in the assumed CMR inputs.

The bound constraints are interpreted as prior information concerning the unknown $\bX$.  Denoting by $\Theta$ the multi-dimensional step function, taking on the value one if all the components of the argument are positive and vanishing otherwise, we write
\[
 \pi_{\rm prior}(\bX)\propto \Theta({\mathsf C}\bX - \bc),
\]
where ``$\propto$'' stands for ``proportional to''. In practice, to obtain a proper prior density, the components need to be restricted further on some wide interval $[-M,M]$, which is in line with the physiological understanding of the system. By Bayes' formula, the posterior density of $\bX$ is then of the form
\[
 \pi(\bX\mid\bR)\propto \Theta({\mathsf C}\bX -\bc){\rm exp}\left(-\frac 12 (\bR-{\mathsf M}\bX)^{\mathsf T}{\mathsf \Sigma}^{-1}(\bR -{\mathsf M}\bX)\right).
\]
In the described framework is it is natural to augment the system with additional information; assuming that we have reasons to believe that a given component $X_j$ is close to a target value, $X_j\approx b_j$, we may simply modify the posterior density by multiplying it with a Gaussian,
\[
 \pi(\bX\mid \bR) \rightarrow \pi(\bX\mid \bR)\times{\rm exp}\left(-\frac 1{2w_j^2}(X_j - b_j)^2\right),
\]
where $w_j^2$ is the variance determining how stringent the condition is believed to be.

Sampling-based strategies for FBA aim at generating a random sample of vectors that satisfy either exactly or approximately the stoichiometric equations with the prescribed bounds. The MCMC methods generate the sample sequentially through a Markov process so that the vectors are distributed according to the posterior density. In particular, if the sample is denoted by
\[
 {\mathscr S} = \big\{X^{(1)},X^{(2)},\ldots, X^{(N)}\big\},
 \]
 the sample allows us to compute approximations for expected values with respect to the posterior density,
 \[
  {\mathsf E}_{\pi(\bX\mid \bR)}\big\{f(\bX)\big\}  = \int f(\bX)\pi(\bX\mid\bR)d\bX \approx \frac 1N\sum_{j=1}^N  f(\bX^{(j)}).
 \]
 For details, see, e.g. \cite{CSbook}.

We apply the sampling algorithm to analyze a spatially distributed system that consists of $N=4$ identical neuron-astrocyte units. In the computation, we assume that each unit has the same household energy need, and that the household energy is distributed between neuron and astrocyte compartments as in the single unit analysis, that is, $H_{\rm n}=H_{\rm a}$. We run three different tests with the following V-cycle configurations: (1) {\em Baseline stimulation}, in which all four units have the same baseline V-cycle activity $V^{(j)} = V/4$; (2) {\em Proximal activation}, in which the unit $n=1$ closest to the capillary has the highest V-cycle activity, $V^{(1)} = 0.9\,V$, while the rest of the activity is distributed among the remaining units,; and (3) {\em Distal activation}, in which 90\% of the V-cycle activity takes place in the unit $n=4$ furthest from the capillary. In each case, we generate a sample that, after cleaving off a burn-in sequence of sample vectors identified as non-representative ones, consist of $N=80\,000$ sample vectors. The burn-in sequence, identified by visual inspection of the time traces of the components, varies from few thousand (uniform and proximal activation) to about 15\,000 (distal activation).

\begin{figure*}
\centerline{\includegraphics[width=4cm]{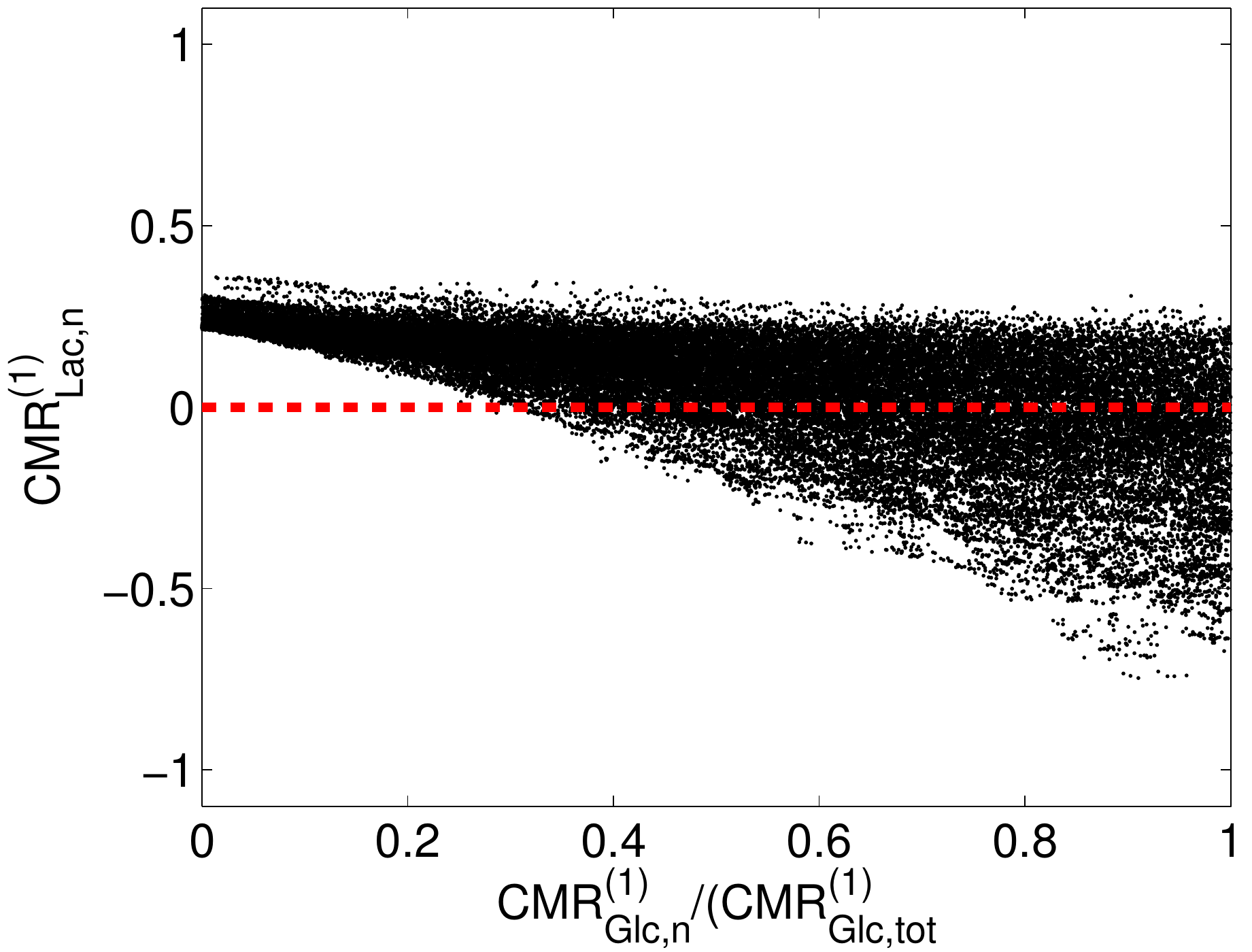}
\includegraphics[width=4cm]{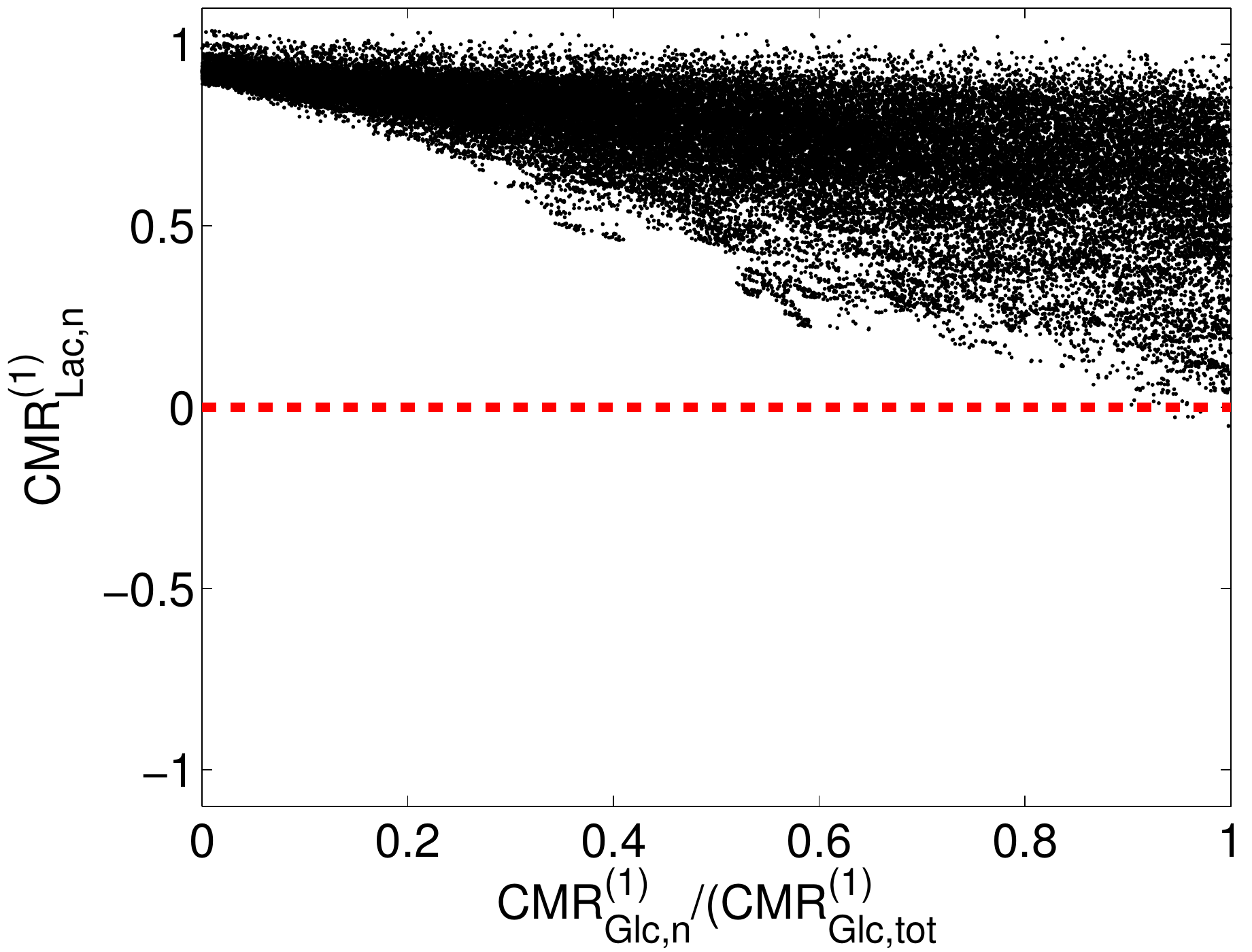}
\includegraphics[width=4cm]{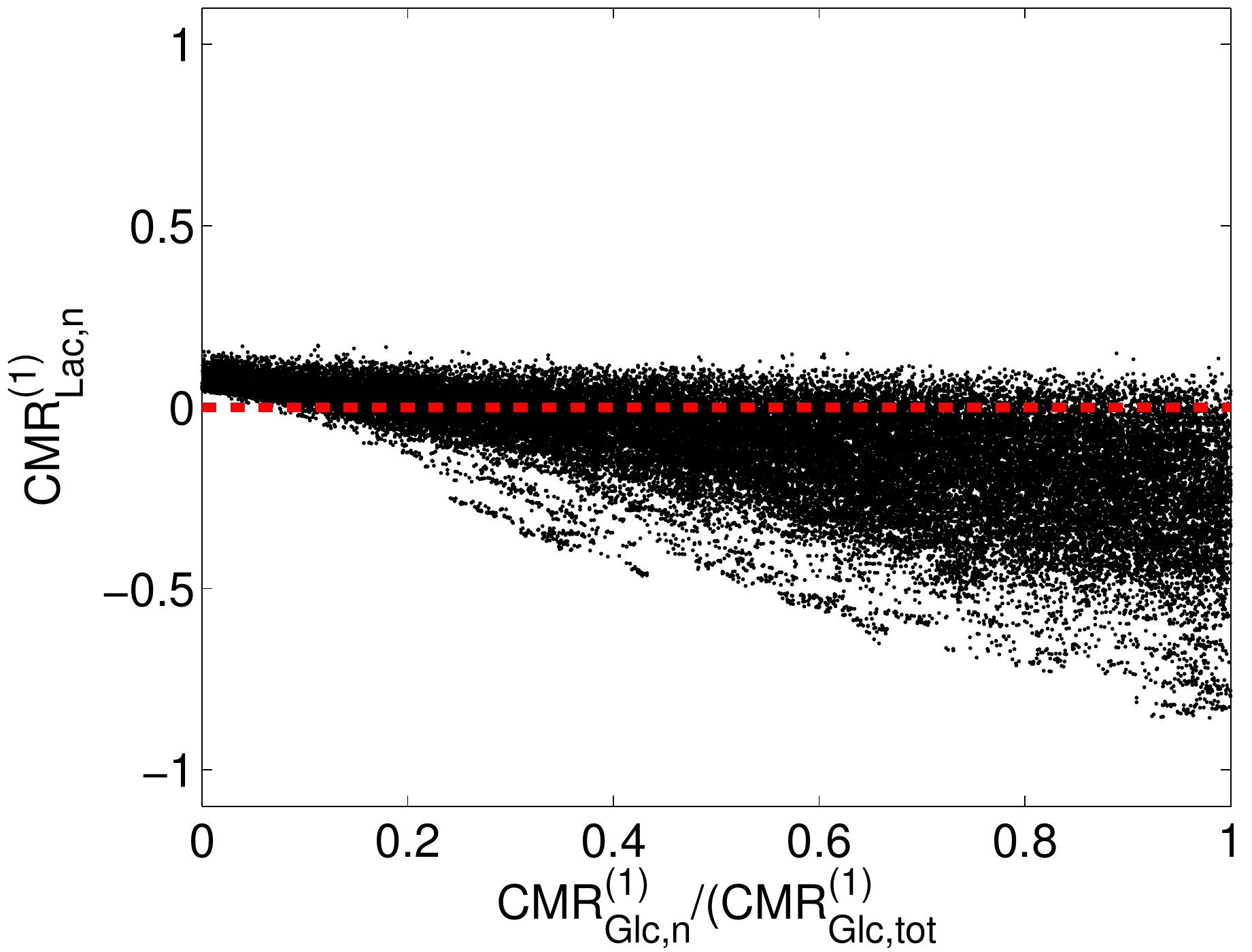}
}
\centerline{\includegraphics[width=4cm]{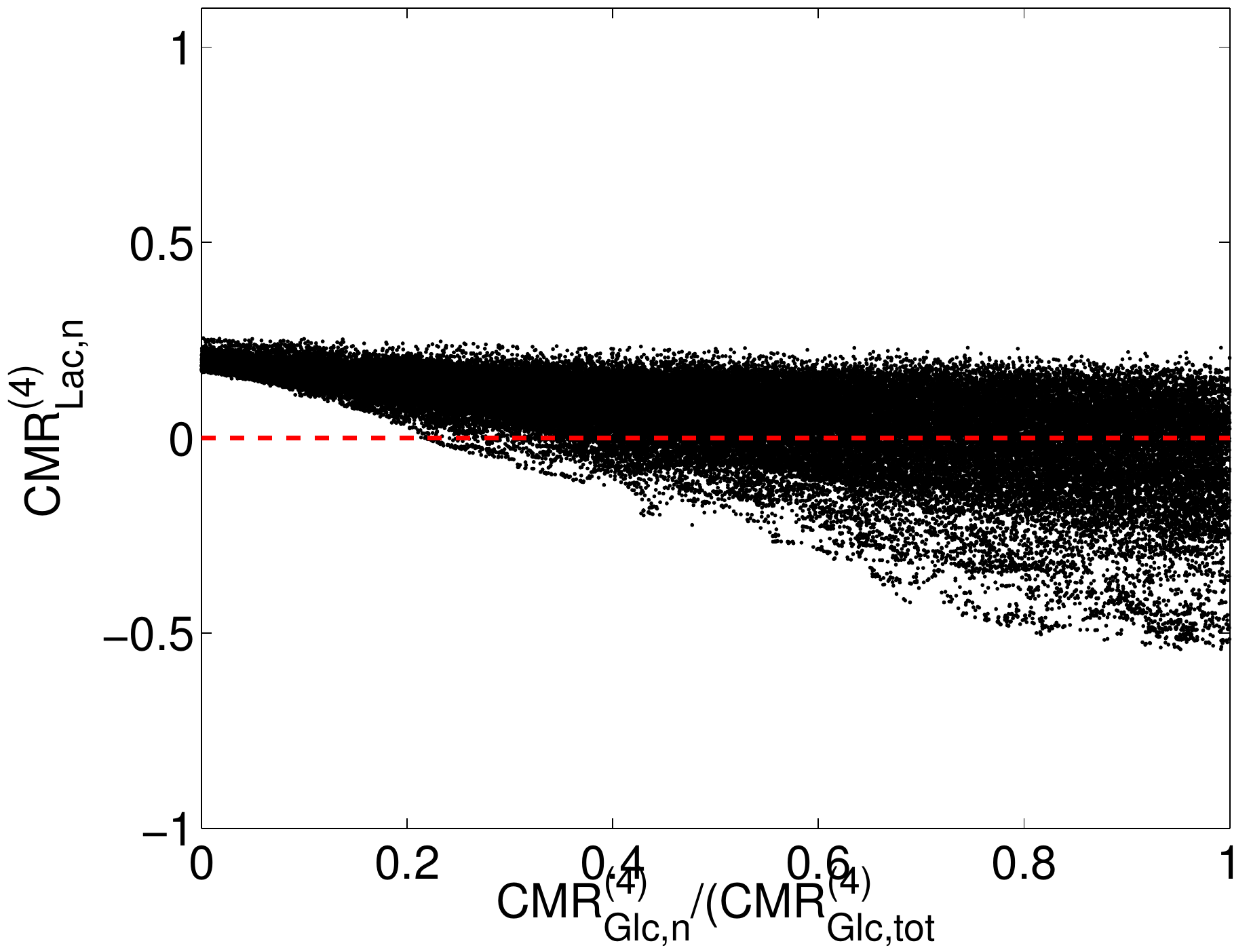}
\includegraphics[width=4cm]{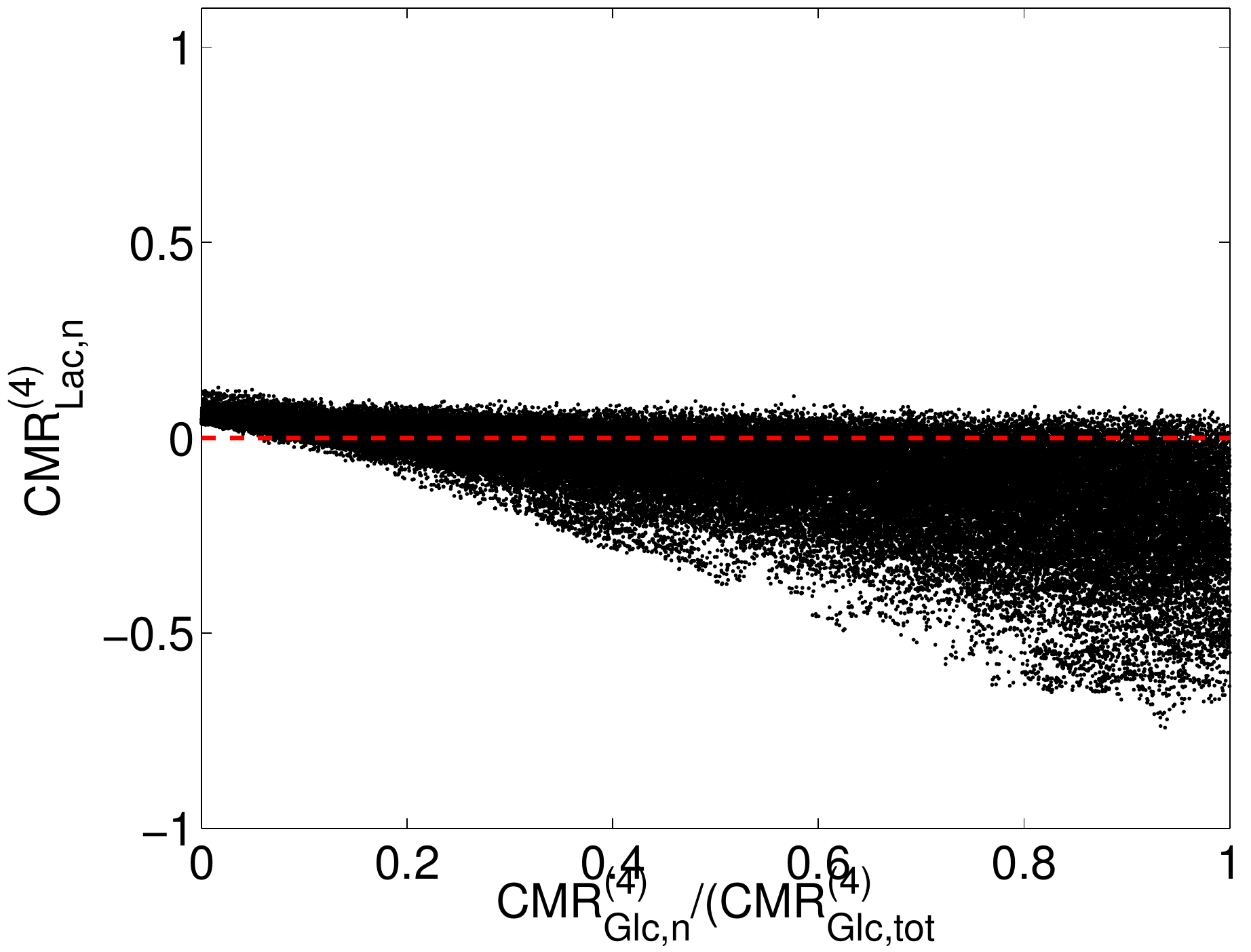}
\includegraphics[width=4cm]{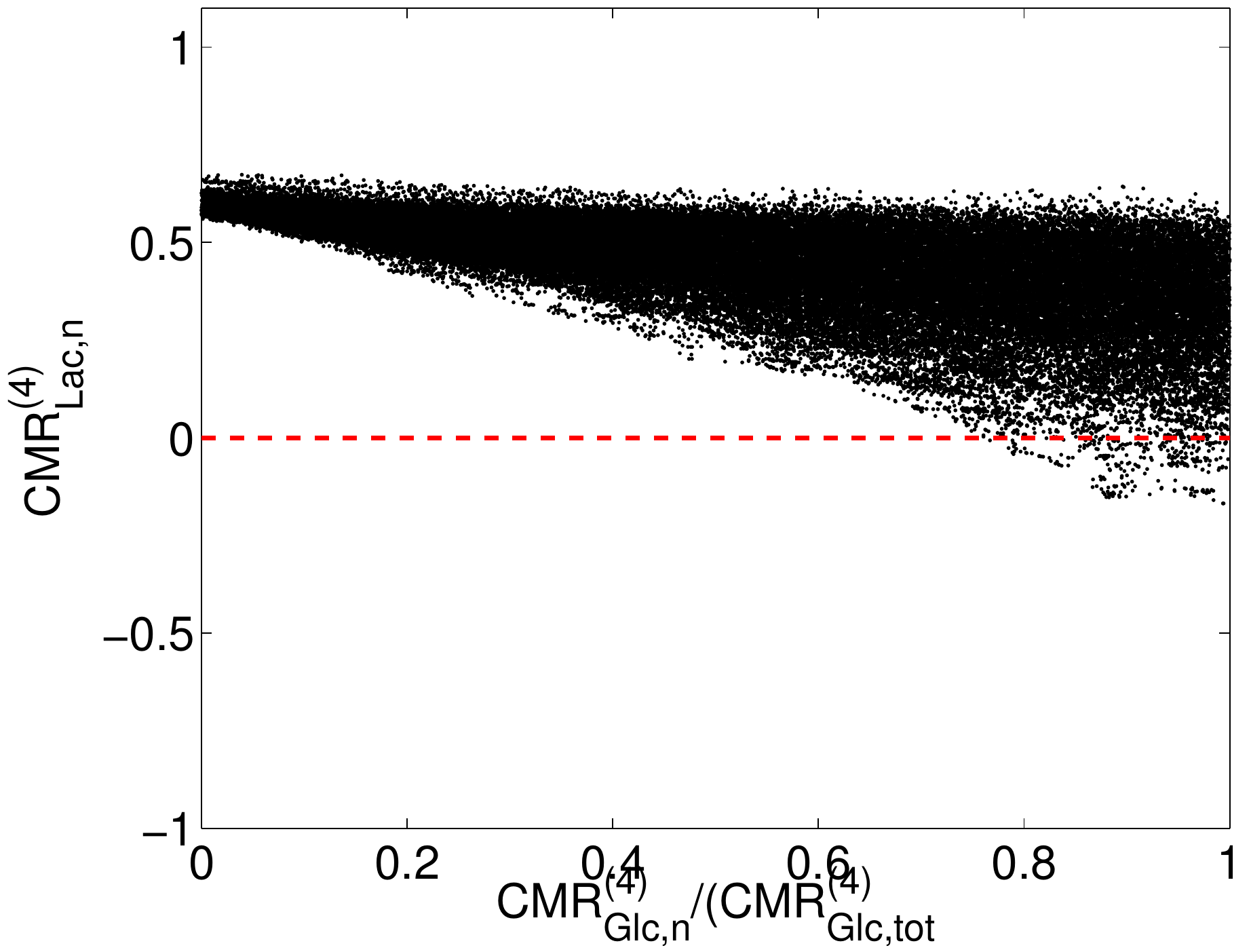}
}
\caption{\label{fig:glucose partitioning} Scatter plots of glucose partitioning versus neuronal lactate uptake in individual units.
The glucose partitioning is expressed in terms of the ratio of glucose uptake by the neuron in the $n$th unit, ${\rm CMR}_{\rm Glc,n}^{(n)}$, and the total glucose uptake of the neuron and astrocyte in the unit, ${\rm CMR}_{\rm Glc,tot}^{(n)}={\rm CMR}_{\rm Glc,n}^{(n)}+{\rm CMR}_{\rm Glc,a}^{(n)}$.  The top row corresponds to the unit nearest to capillary, and the bottom row to the most distant unit from the capillary. On the left, the V-cycle activity is uniformly divided among the units, in the middle, the most active unit is the first one, and on the right, most of the V-cycle activity is in the last unit. The horizontal line indicates the zero level, in which the neuron switches from lactate production (``$-$'') to lactate oxidation (``$+$''). Observe that unlike in a single unit lumped model with only two degrees of freedom, the lactate traffic no longer depends solely on the glucose partitioning inside the unit, since the stoichiometry allows a wide spectrum of different configurations, in which the lactate may be produced or taken up  by other units. The figure clearly demonstrates that the neuron in the active unit is oxidizing lactate, while the in the units of low activity, the status is not certain.}
\end{figure*}

We start by investigating whether the increased number of degrees of freedom changes the conclusion obtained with the lumped model that the lactate trafficking between astrocyte and neuron is determined by the glucose partitioning. In Figure~\ref{fig:glucose partitioning}, we plot the glucose partitioning between neuron and astrocyte within a single unit, versus the neuronal lactate uptake in that unit. The scatter plots clearly show that the neuron in the most active unit is with a high probability a lactate user, as the lactate flux for most realizations is positive, while in the less active units, the flux is negative, indicating that the neuron may be a lactate producer. Next we observe that the lactate flux is no longer predictable by the glucose partitioning. In particular,  when the ratio ${\rm CMR}_{\rm Glc,n}/{\rm CMR}_{\rm Glc,tot}$ approaches one, indicating that the neuron in the unit uptakes all glucose, the variability of the lactate flux becomes significant. In the case of proximal and distal activation, the flux into the active neuron remains predominantly positive, while in the uniform activation scheme, even the direction of the lactate traffic is undetermined. The explanation for this behavior is that the other units, neurons in them included, may either produce or uptake lactate, and the local relation between neuron and astrocyte becomes more obfuscated.

\begin{figure*}
\centerline{\includegraphics[width=4cm]{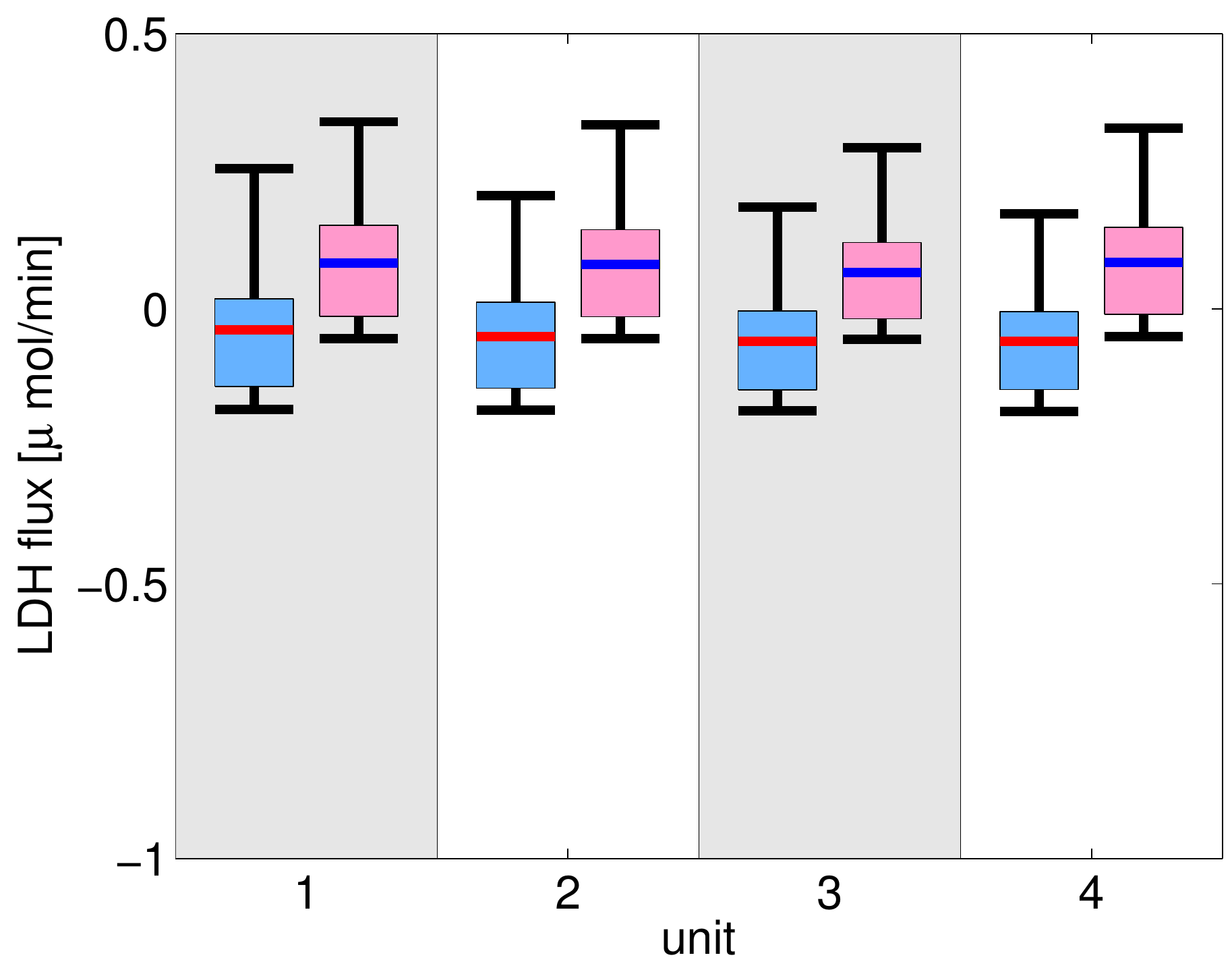}
\includegraphics[width=4cm]{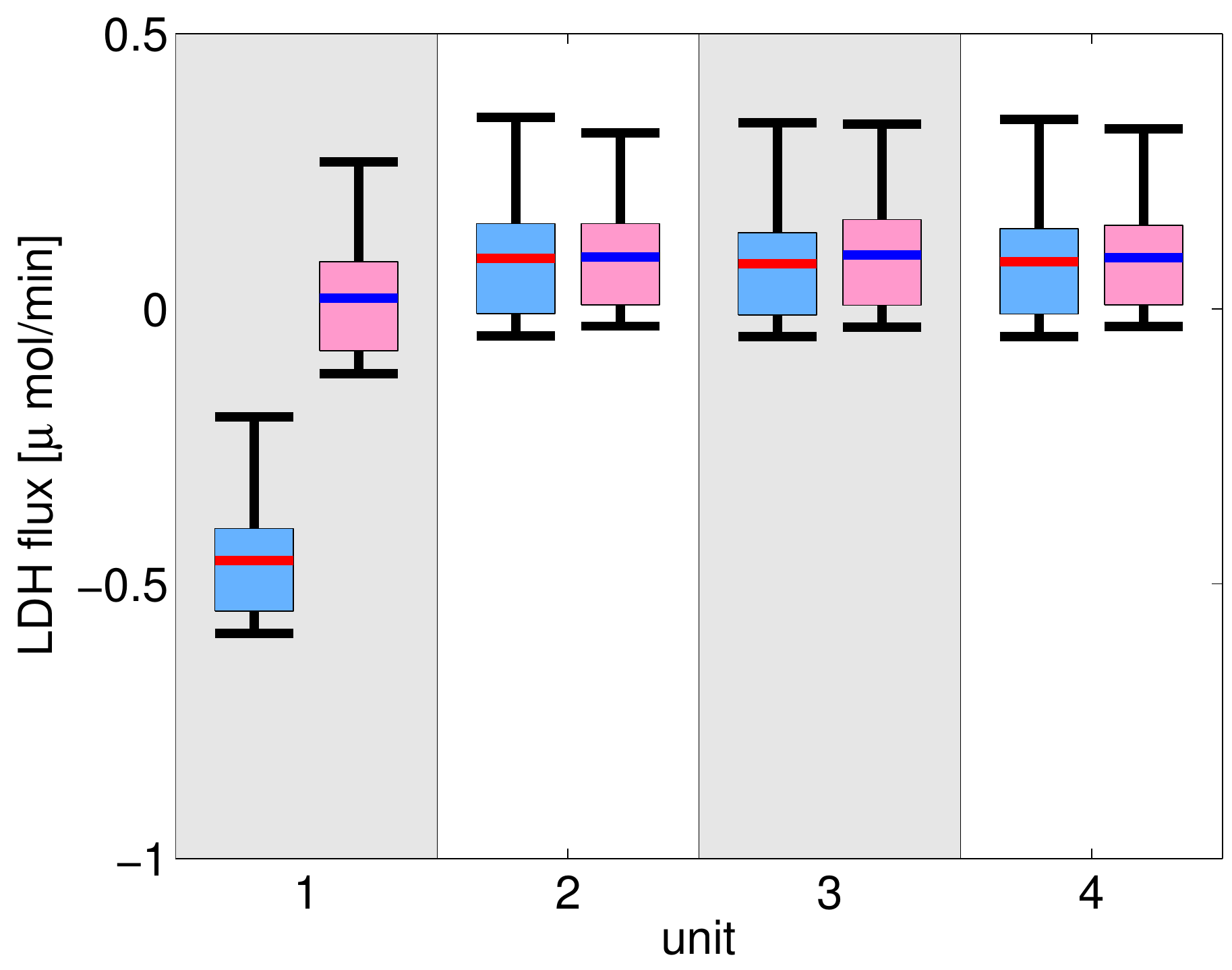}
\includegraphics[width=4cm]{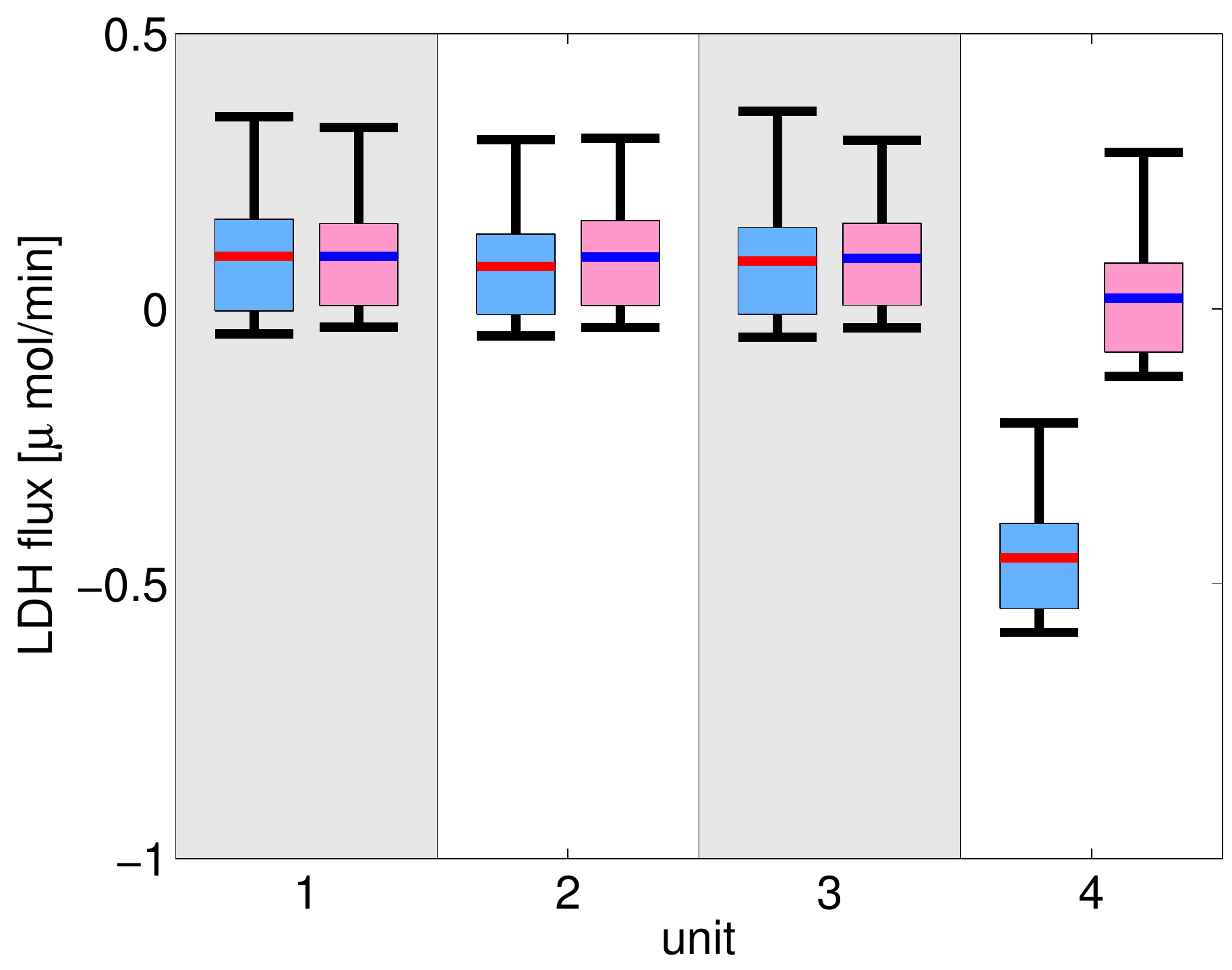}
}
\caption{\label{fig:LDH flux}  LDH reaction fluxes in neuron and astrocyte in the five units. The blue boxes refer to neurons and the pink ones to astrocytes. On the left, the V-cycle activity is uniformly distributed among the units, in the middle, the active unit is $n=1$ next to the capillary, on the right at the farthest unit $n=4$. The boxes indicate 50\% belief, and the whiskers 90\% belief intervals. Observe that in the uniform activation, the neurons are predominantly oxidizing lactate (LDH negative) while the astrocytes produce lactate (LDH positive). In the distal and proximal activations, independently of the location of the high activity, the neuron oxidizes lactate only in the active unit, while in the other units, astrocytes and neurons have a very similar role as lactate producers.}
\end{figure*}

To shed some light on which compartments produce or uptake lactate, we plot the distribution of the LDH activities in each compartment. Figure~\ref{fig:LDH flux} summarizes the distributions as box plots, indicating the mean value of the flux over the sample as well as the 50\% and 90\% belief intervals computed from the sample. The plot shows that in the case of uniform activation, the neurons favor negative LDH, or lactate oxidation, while in astrocytes, reductive LDH is favored.
 Interestingly, in the proximal and distal activation configuration,  the only compartment oxidizing lactate is the neuron in the most active unit, while all the other units are lactate producers. In particular, the difference between neuron and astrocyte in a non-active unit is minimal.

\begin{figure*}
\centerline{\includegraphics[width=4cm]{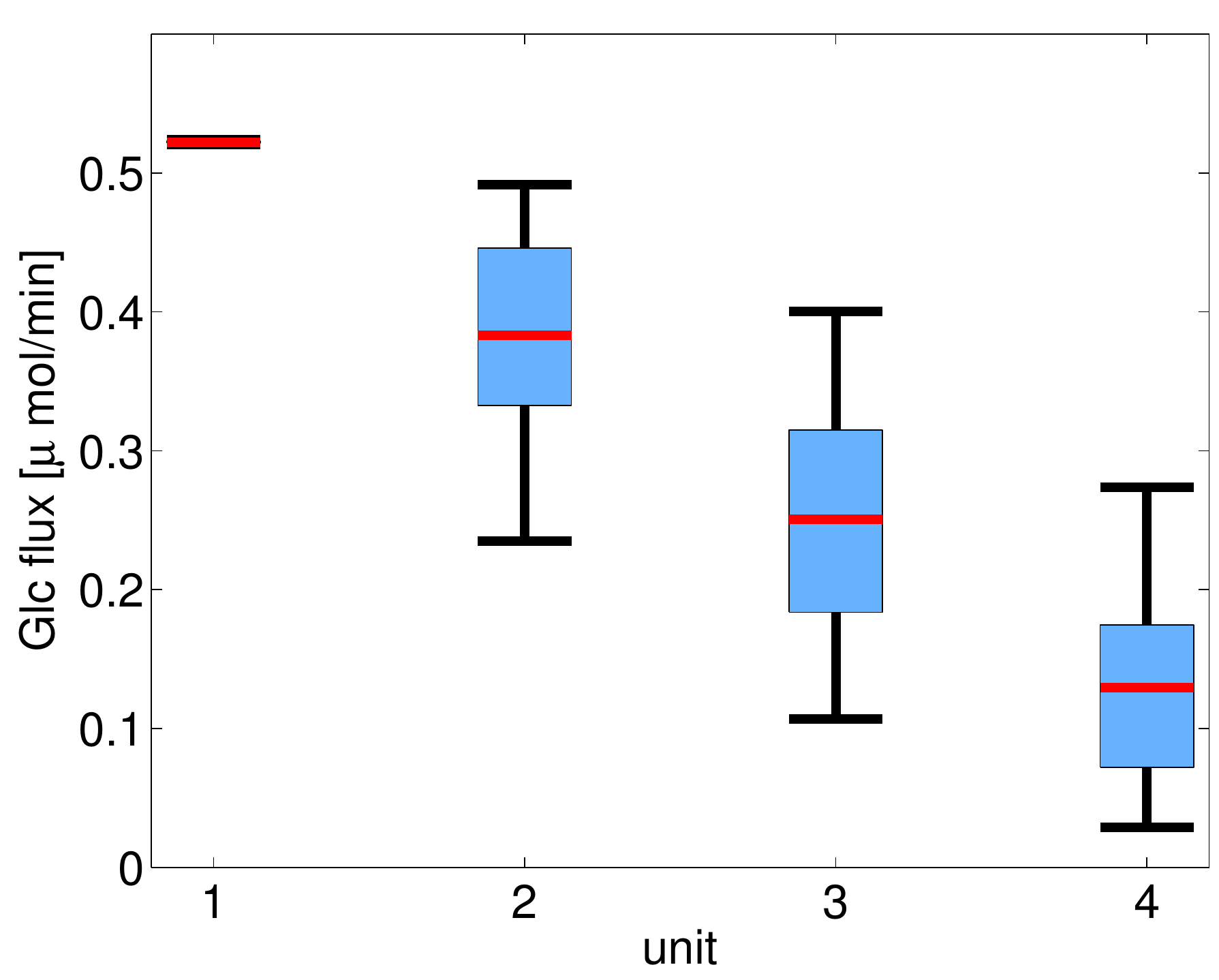}
\includegraphics[width=4cm]{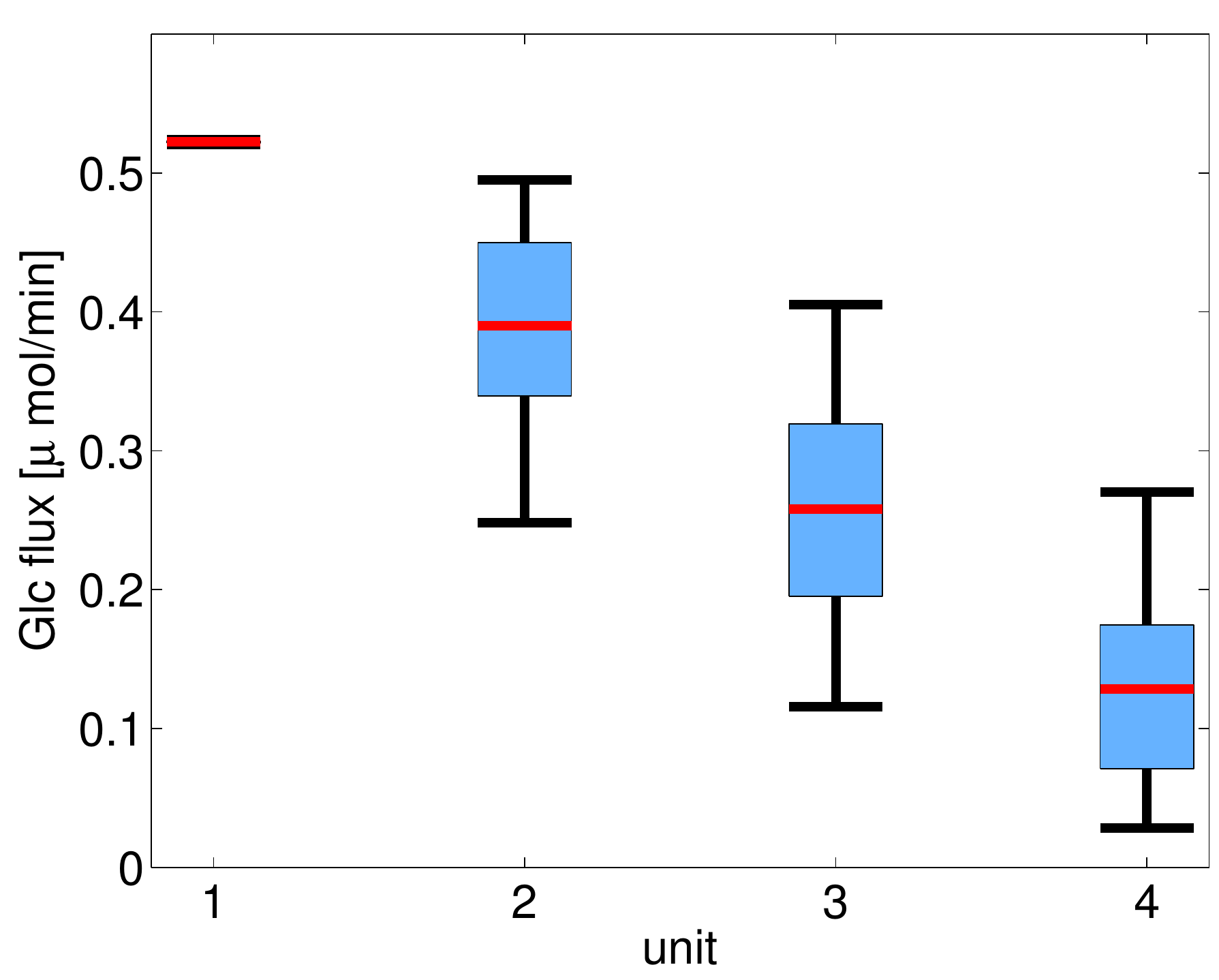}
\includegraphics[width=4cm]{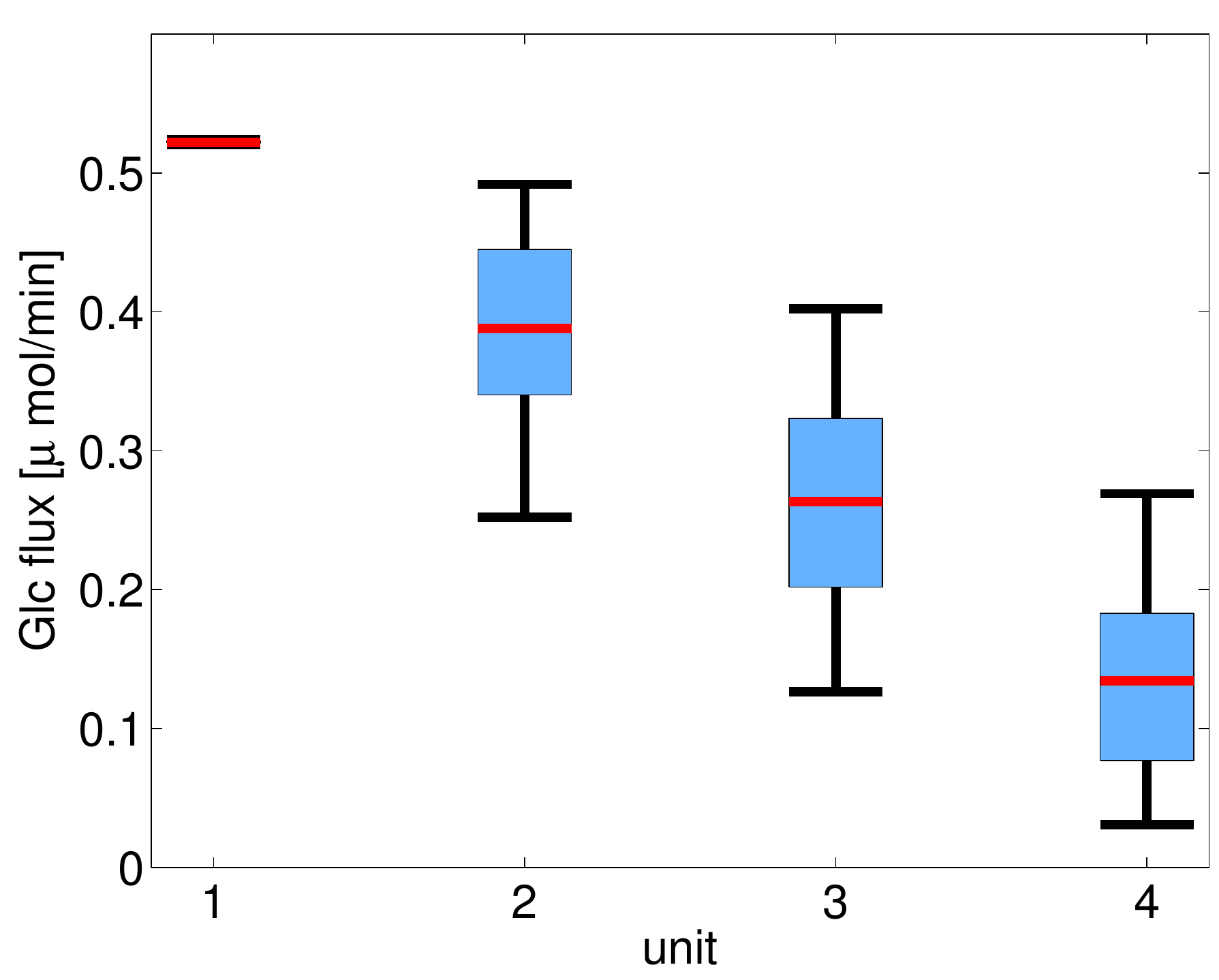}
}
\centerline{\includegraphics[width=4cm]{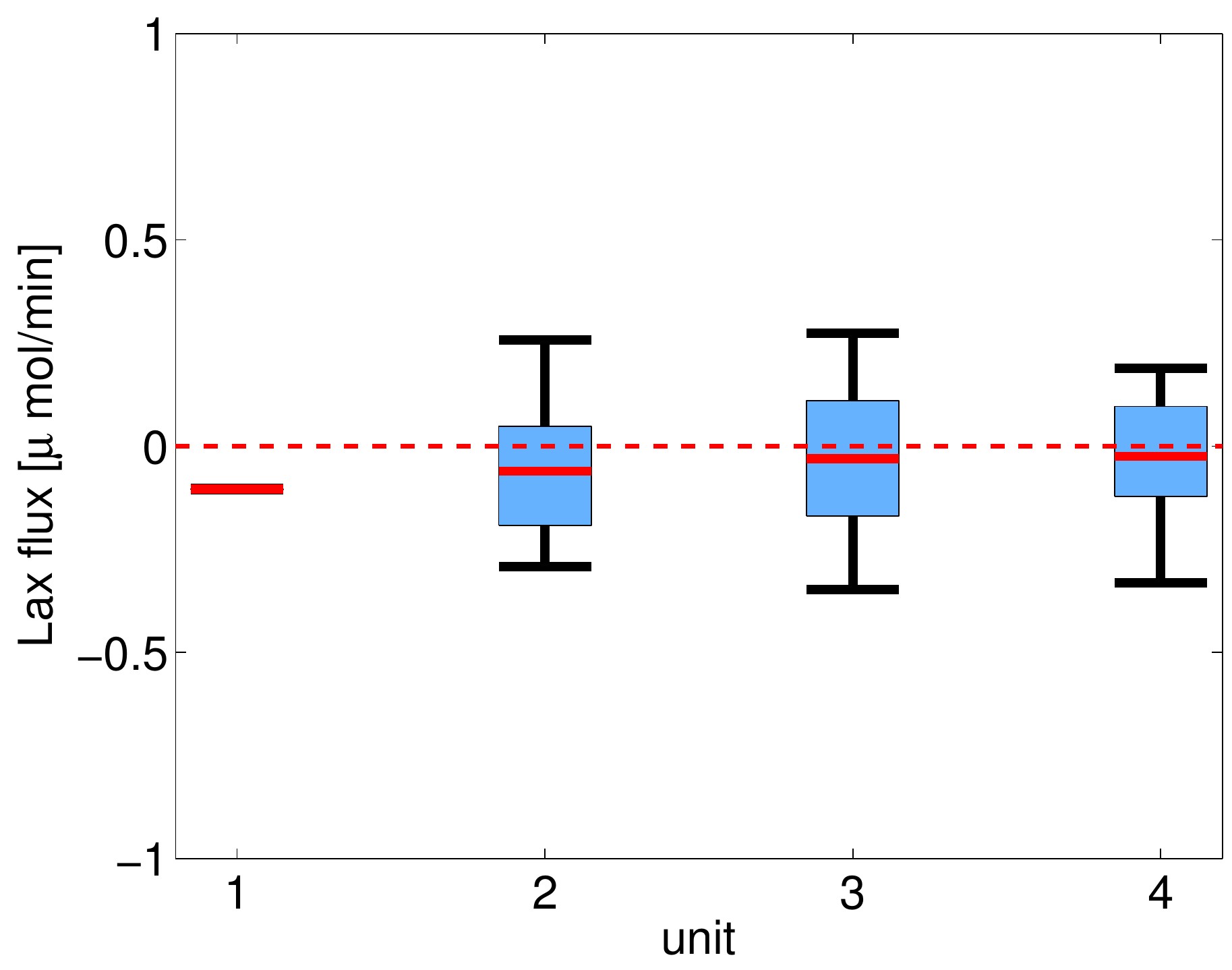}
\includegraphics[width=4cm]{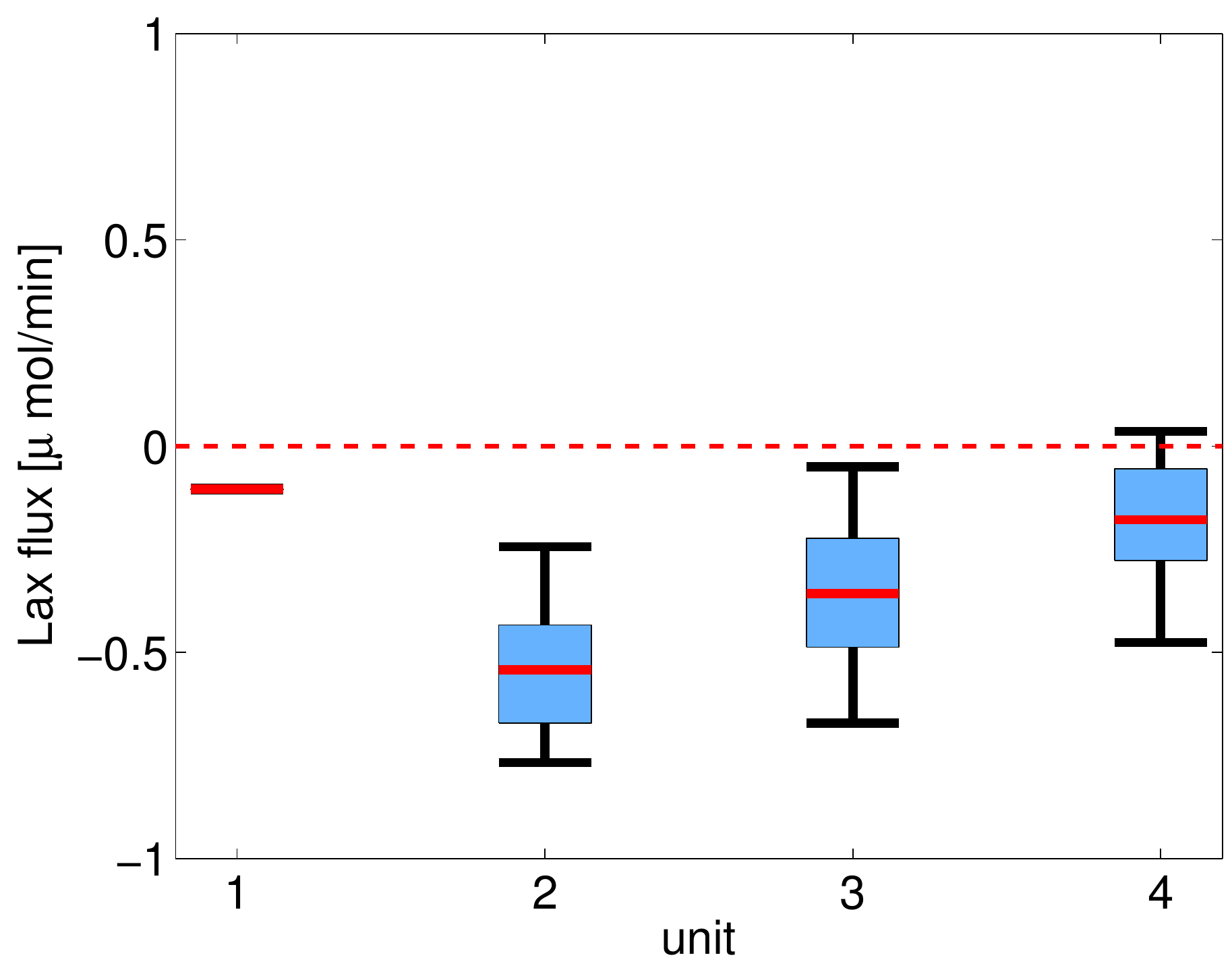}
\includegraphics[width=4cm]{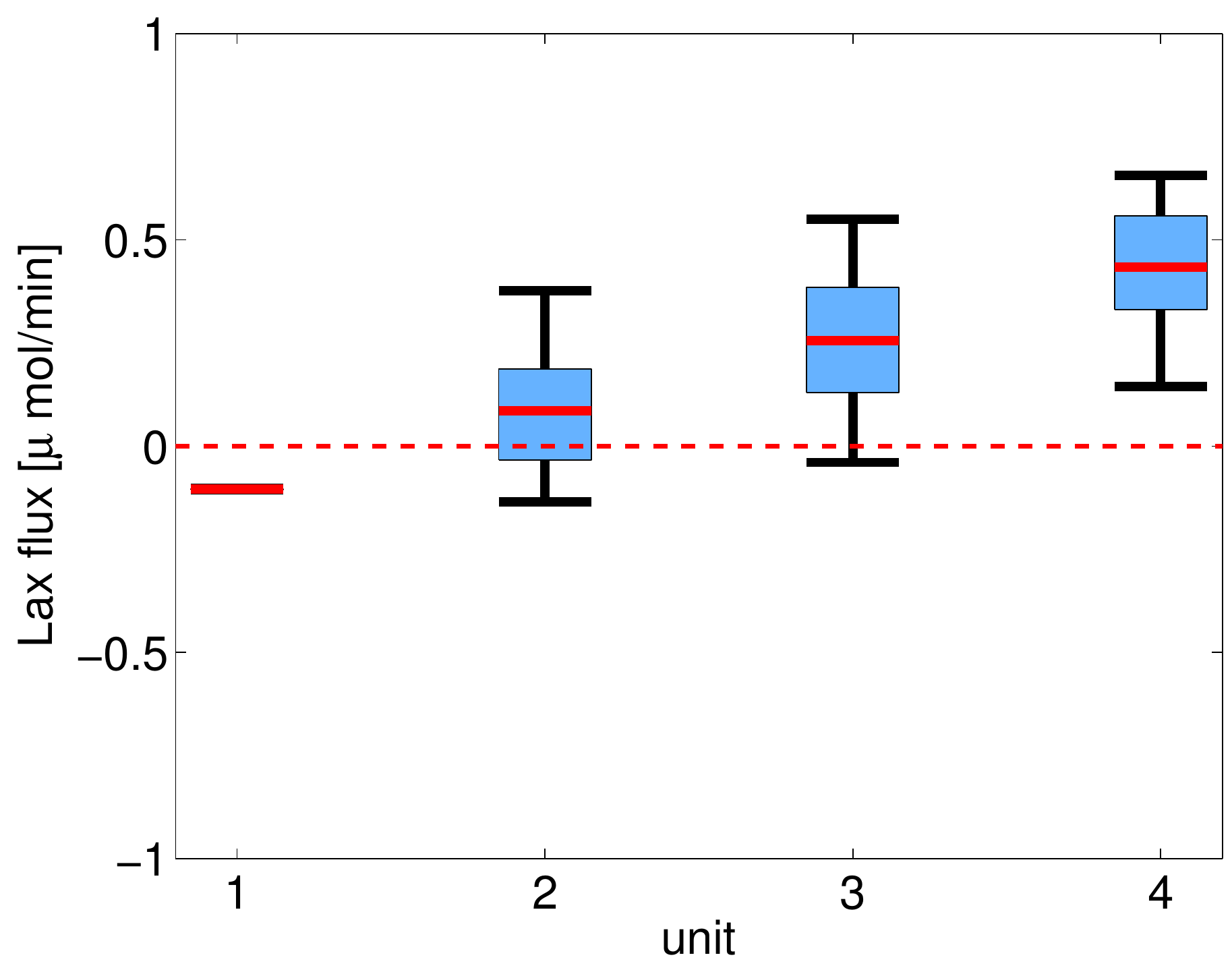}
}
\centerline{\includegraphics[width=4cm]{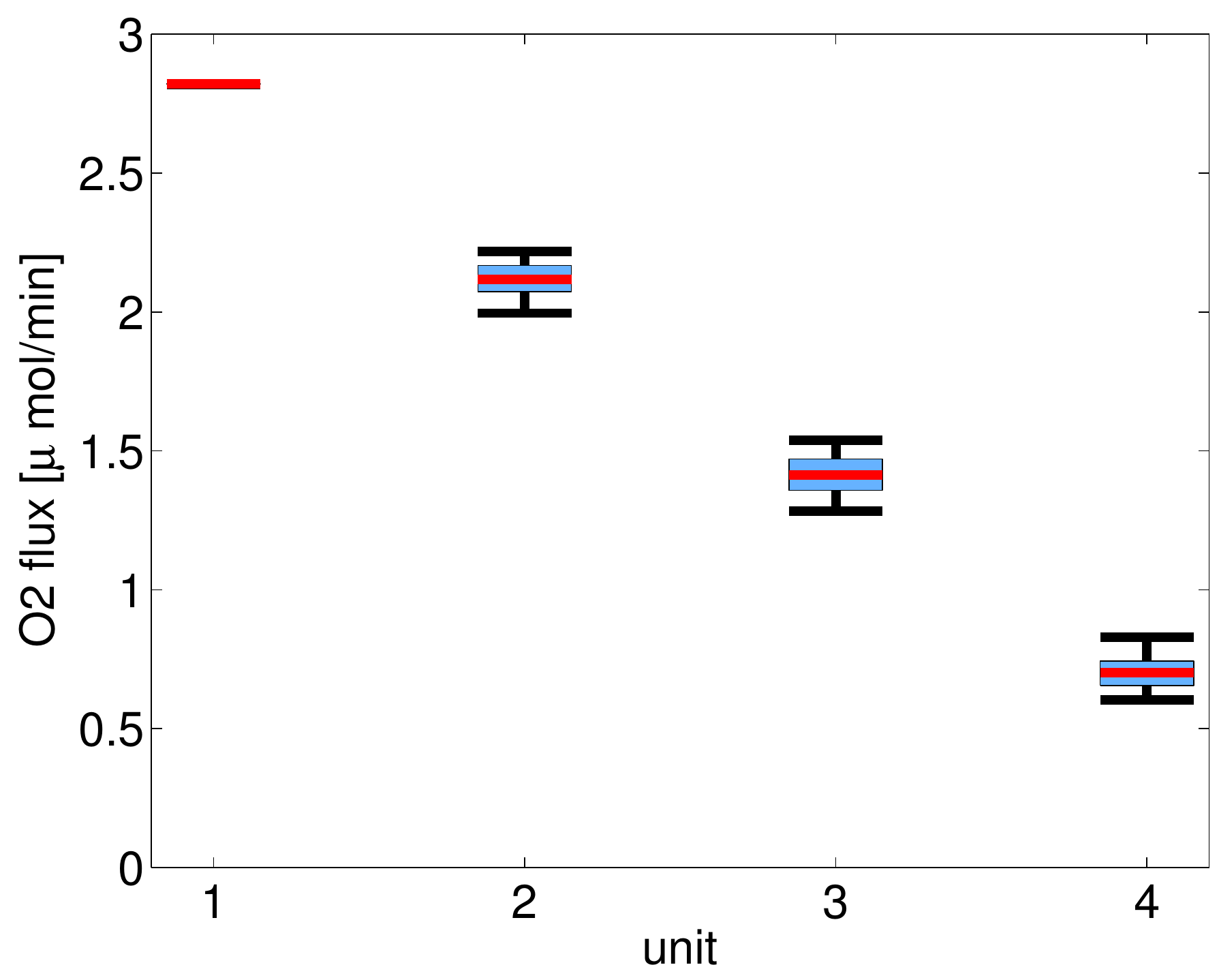}
\includegraphics[width=4cm]{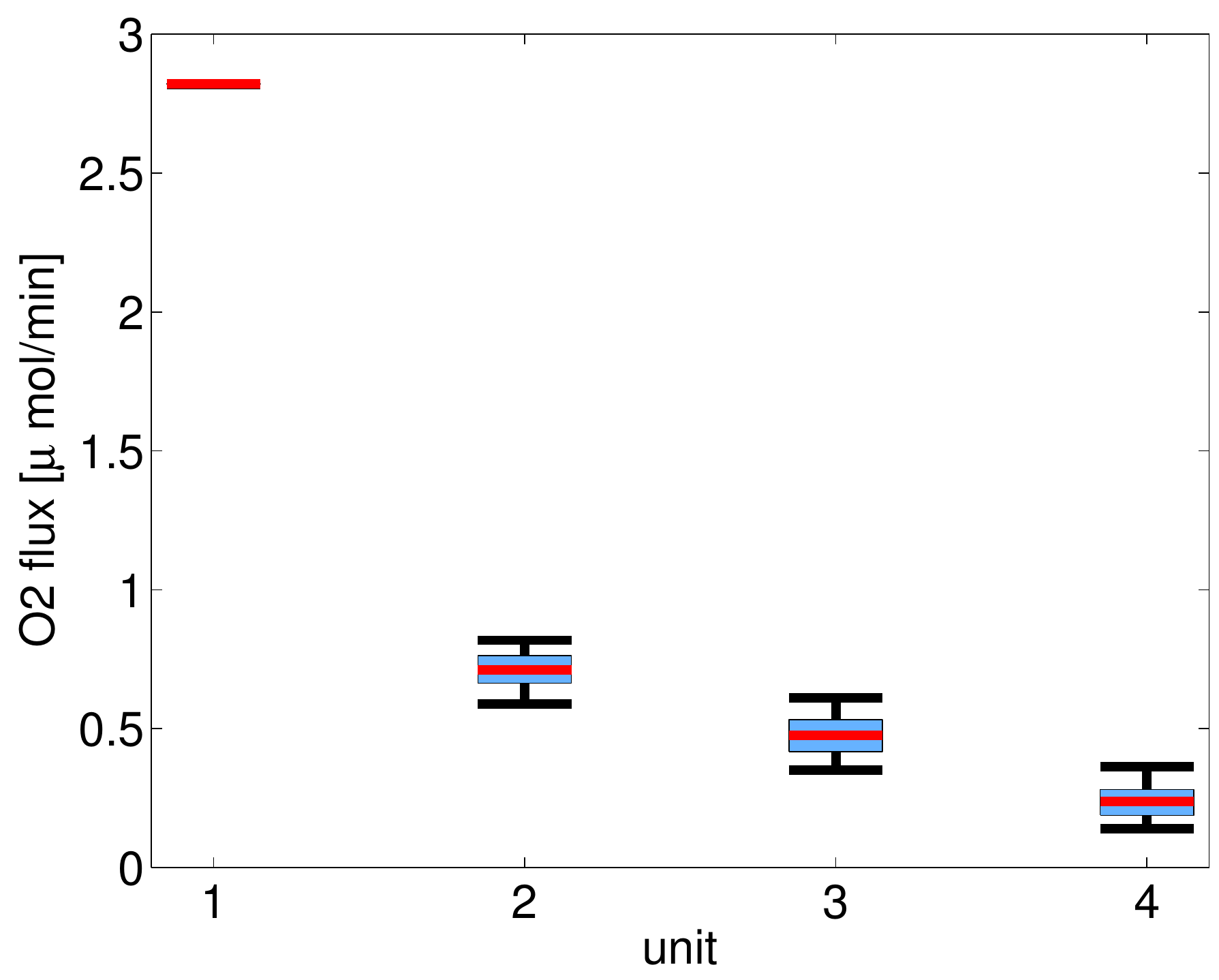}
\includegraphics[width=4cm]{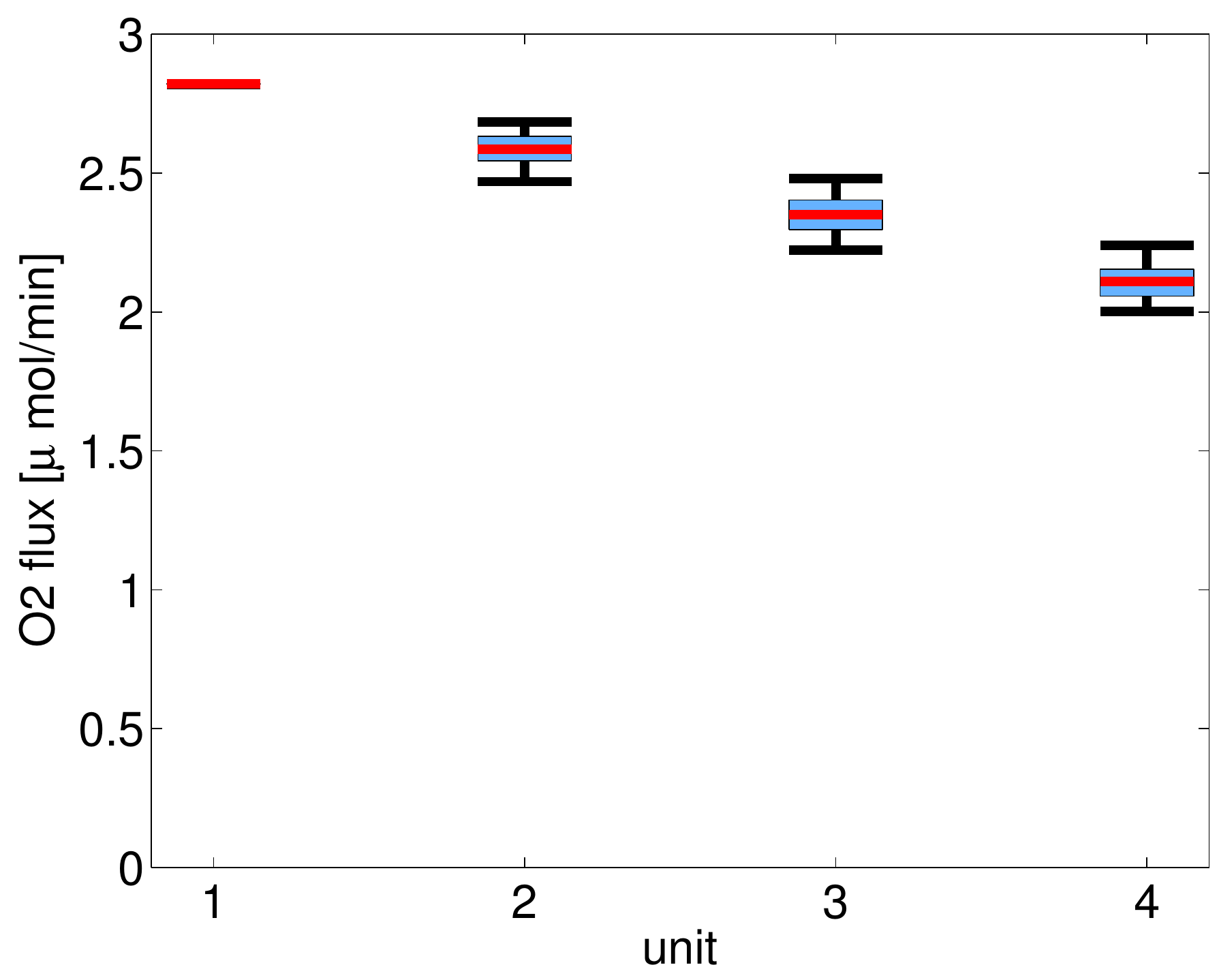}
}
\caption{\label{fig:diffusion fluxes}The mean diffusion fluxes with 50\% (box) and 90\% (whiskers) uncertainty intervals. On the left, the activation is uniform, in the middle, the unit with the highest V-cycle activity is closest to the capillary (Unit 1), while on the right, most of the V-cycle activity takes place in the most distant unit (Unit 4). The glucose diffusion (top row) is not affected by the location of the activation, while the lactate (middle) and oxygen fluxes (bottom) are significantly different as the location of the activation changes: In the proximal activation, the units further away from the capillary receive very little oxygen and therefore produce lactate which flows towards the first unit, where the neuron is a net lactate oxidizer. In the distal activation, oxygen flux through the system is high, and lactate is flowing towards the active unit where the neuron takes it up and oxidizes it.}
\end{figure*}

Finally, we plot the distributions of the diffusion fluxes of glucose, lactate and oxygen, see Figure~\ref{fig:diffusion fluxes}. The glucose diffusion patterns are almost identical regardless of the activation pattern, however, the lactate and oxygen fluxes differ significantly. In the proximal excitation, the neuron in the first unit is highly  oxidative, with the result that the oxygen flux decreases significantly deeper down in the tissue. The compartments with low oxygen availability run non-oxidative glycolysis, producing lactate that diffuses upwards towards the capillary (negative diffusion flux). In the distal activation, where the highly oxidative neuron is in the last unit, the oxygen diffusion flux is high. However, despite the oxygen availability, the units closer to the capillary run non-oxidative glycolysis, thus producing lactate which diffuses towards the active neuron that uptakes lactate.

\section{Conclusions}

An analysis of the energetic needs of a lumped neuron-astrocyte neurovascular unit with a stoichiometric model which can be treated analytically provides an explicit relation between the oxidative glucose metabolism, the glutamatergic neurotransmitter activity and household energy of the cells. This formula can be used, in particular, to find an estimate for the energetic cost of running the V-cycle.
Using the experimental results of the oxidative glucose metabolism and neurotransmitter cycling in the human brain \cite{Lebon2002}, the model suggests that for each glutamate passing through the V-cycle,  31 ATP in neuron and 5 ATP in astrocyte need to be produced to cover the energetic cost arising from signal propagation and transmission, as well as the household maintenance. Compared to the estimate in  \cite{Howarth2012}, 57 ATP in total per one glutamate, these numbers are somewhat lower than in the cited article, which may be partly attributed to the difference between rodent and human brain metabolism.

The model derived in this article is also used to estimate the traffic of lactate between astrocyte and neuron as a function of the parameters encoding the degrees of freedom of the system. The analysis corroborates the previously published results, according to which the lactate traffic at steady state depends primarily on the glucose partitioning between neuron and astrocyte. However, when carrying out the analysis in a spatially distributed variant of the model which accounts for metabolite diffusion in ECS, the lactate trafficking stops being a local phenomenon, and it is no longer clear if its direction is solely a function of glucose partitioning. To shed more light on this question, an analysis with a refined spatially distributed model (\cite{Calvetti2014}) is needed, and the findings will need be confirmed with experimental validation.

\end{document}